\documentclass[fleqn]{report}

\let\Bar\overline 

\usepackage[psamsfonts]{amssymb} 

\usepackage[dvips]{graphicx}
  
 \newtheorem{theorem}{Theorem} 
 \newtheorem{definition}[theorem]{Definition}
 \newtheorem{corollary}[theorem]{Corollary}
 \newtheorem{lemma}[theorem]{Lemma}
 \newtheorem{Lem}[theorem]{Lemma}
 \newenvironment{proof}{\par\noindent{\it Proof.\ }}{\par}

\def\itx#1{{\it#1}}
\def\inx#1{#1}

\def\Qm#1{\M_{#1}}\def\Cl#1{\C_{#1}}
\def\captyc{C_c}\def\captyq{C_q}
\def\fidel{{\cal F}}

\def\pp#1/#2/{{\mathbb P}(#1\mid#2)}
\def\A{{\cal A}}\def\B{{\cal B}}\def\C{{\cal C}}\def\Hh{{\cal H}}
\def\K{{\cal K}}\def\M{{\cal M}}

\def\Cx{{\mathbb C}}
\def\err{\Delta}\def\capty{C}
\def\idty{{\rm 1\mkern -5.4mu I}}
\def\id{{\rm id}}
\def\tr{{\mathop{\rm tr}\nolimits}}
\def\ketbra#1#2{\vert#1\rangle\langle#2\vert}
\def\bra#1>{\langle#1\rangle}
\def\ket#1{\vert#1\rangle}
\def\abs#1{\vert#1\vert}
\def\norm#1{\Vert#1\Vert}
\def\cbnorm#1{\norm{#1}_{\rm cb}}

\begin{document} 

 
\begin{titlepage}
 \null\vfil
  \vskip 30pt
  \begin{center}%
    {\LARGE Quantum Information Theory -- an Invitation \par}%
    \vskip 3em%
    {\large \lineskip .75em%
      R.F. Werner$^\ddagger$\vskip 1.5em}\par
      Institut f\"ur Mathematische Physik, TU Braunschweig\\
      Mendelssohnstr. 3 / 38106 Braunschweig / Germany \\
    {\large \vskip 3em October 30, 2000 \par}%
  \end{center}
  \vskip 60pt      
    
\noindent This text is part of a volume entitled ``{\it Quantum 
information --- an introduction to basic theoretical concepts and 
experiments}'', to be published in {\it Springer Tracts in Modern 
Physics}. Authors will be G.~Alber, T.~Beth, M., P., and R.  
Horodecki. M. R\"otteler, H. Weinfurter, R.F. Werner, and A. 
Zeilinger. Articles: 

\vskip12pt 

\begin{tabular}{ll}
  \bf G. Alber&From the foundations of quantum theory\\&
        to quantum technology - an introduction
\\ \noalign{\vskip5pt} 
  \bf R.F. Werner&This article\\ \noalign{\vskip5pt}
  \bf H. Weinfurter, A. Zeilinger&Quantum communication \\ \noalign{\vskip5pt}
  \bf Th. Beth, M. R\"otteler&Quantum algorithms: applicable \\
                        &\quad algebra und quantum physics\\ \noalign{\vskip5pt}
  \bf M., P., and R.~Horodecki&Mixed-state entanglement \\ 
                     &\quad and quantum communication\\ \\
  &Joint index\\
  &Joint list of references
\end{tabular} \vskip 20pt 

\noindent References to ``articles in this book'' will be to these 
articles.

\vfill\noindent{\large$^\ddagger$}%
 Email: {\tt r.werner@tu-bs.de}\newline
 Web:  {\tt http://www.imaph.tu-bs.de/qi} 
 
\end{titlepage} 

 \tableofcontents

\chapter{Introduction}

Quantum Information and Quantum Computers have received a lot of
public attention recently. Quantum Computers have been advertised
as a kind of warp drive for computing, and indeed the promise of
the algorithms by Shor and Grover is to perform computations which
are extremely hard or even provably impossible on any merely
``classical'' computer. On the experimental side, perhaps the most
remarkable feat of Quantum Information processing was the
realization of ``quantum teleportation'', which once again has
science fiction overtones.

In some sense these miracles are an extension of the Strangeness
of Quantum Mechanics -- those unresolved questions in the
foundations of quantum mechanics, which most physicists know
about, but few try to tackle directly in their research. However,
trying to build an explanation of Quantum Information on the
foundations literature is more likely to mystify than to clarify.
It would also give the wrong idea of how discussions in this new
field are conducted. Because, just like physicists of widely
differing convictions on foundational matters can usually agree
quite easily on what the predictions of quantum mechanics are in a
particular experimental setup, researchers in Quantum Information
can agree on whether a device should work, no matter what they may
think about the deeper meaning of the wave function. For example,
one of the founders of the field is an outspoken proponent of the
\inx{Many-Worlds interpretation} of quantum mechanics (which I
personally find useless and bizarre). But whatever the intuitions
leading him to his discoveries about quantum computing may have
been, these discoveries make sense in every other interpretation.

In this article I will give an account of the basic concepts of
Quantum Information Theory, staying as much as possible in the
area of general agreement. So in order to enter this new field,
plain quantum mechanics is enough, and no new, perhaps obscure,
views are needed. There is, of course, a characteristic shift in
emphasis expressed by the word ``information'', and we will have
to explore the consequences of this shift.

The article is divided in two parts. The first (up to Section~5)
is mostly in plain English, centered around the exploration of
what can or cannot be done with quantum systems as information
carriers. The second part, Section~6, then gives a description of
the mathematical structures and some of the tools needed to
develop the theory.

\chapter{What is Quantum Information?}

Let us start with a preliminary definition:

\begin{quote} {\it \inx{Quantum Information} is that kind of information, which
is carried by quantum systems from the preparation device to the
measuring apparatus in a quantum mechanical experiment.}
\end{quote}

\noindent So a ``transmitter'' of quantum information is nothing
but a device preparing quantum particles, and a ``receiver'' is
just a measuring device. Of course, this is not saying much. But
even so, it is a strange statement from the point of view of
classical information theory: in that theory one usually does not
care about the physical carrier of information, or else one would
also have to distinguish ``electrodynamical information'',
``printed information'', ``magnetic information'', and many more.
In fact, the success of (classical) information theory depends
largely on abstracting from the physical carrier, and going
instead for the general principles underlying any information
exchange. So why should ``quantum information'' be any different?

A moment's reflection makes clear why the abstraction from the
\inx{physical carrier of information} leads to a successful theory: the
reason is that it is so easy to convert information between all
those carriers. The conversion from bytes on a hard disk, to
currents in a chip, to signals on a net cable, to radio waves via
satellite, and maybe finally to an image on a computer screen in
another continent all happen essentially without loss, and if
there are losses, they are well understood, and it is known how to
correct for them. Therefore the crucial question is: can ``quantum
information'' in the above loose sense also be converted to those
standard classical kinds of information, and back, without loss?
Or else: are there fundamental limitations to such a translation,
and is quantum information hence really a {\it new} kind of
information?

This book would not have been written if the answer to the last
question were not affirmative: quantum information is indeed a new
kind. But to make this precise, let us see what would be required
of a successful translation. Let us begin with the conversion of
quantum information to classical information: a device for this
conversion would take a quantum system and produce as its output
some classical information. This is nothing but a complicated way
of saying ``measurement''. The reverse translation, from classical
to quantum information, obviously involves some preparation of
quantum systems. The classical input to such a device is used to
control settings of this preparing device, and any dependence of
the preparation process on classical information is admissible.
There are two kinds of devices we can combine from these two
elements. Let us first consider a device going from classical to
quantum to classical information. This is a rather commonplace
operation. For example, one can encode one classical bit on the
polarization degree of freedom of a photon (clearly a quantum
system), by  choosing one of two orthogonal polarizations for the
photon, depending on the value of the classical bit. The readout
is done by a photomultiplier combined with a polarization filter
in one of these directions. In principle, this allows a perfect
transmission.  In some sense every transmission of classical
information is of this kind, because every physical system
ultimately obeys the laws of quantum mechanics, even if we can
often disregard this fact and treat it classically. Hence
classical information can be translated into quantum (and back).

\begin{figure}[htbp]
  \begin{center}
    \includegraphics[scale=1]{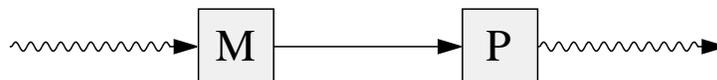}
    \caption{Classical Teleportation. Here and in the following diagrams,
    a wavy arrow stands for quantum systems, and a straight one for
    the flow of classical information.}
    \label{fig:telepo}
  \end{center}
\end{figure}

But what about the converse? This hypothetical (and in fact,
impossible) process has come to be known as \itx{classical
teleportation} (see Figure~\ref{fig:telepo}). It would involve a
measuring device M, operating on some input quantum systems. The
measuring results are subsequently fed into a preparing device P,
which produces the final output of the combined device. The task
is to set things up such that the outputs of the combined device
are indistinguishable from the quantum inputs.
Of course, we have to say precisely, what ``indistinguishable''
should mean. Clearly, this cannot mean that ``the same'' system
comes out at the other end. In the classical case this is not
demanded either. What can only be meant in quantum mechanics is
that {\it no \inx{statistical test} will see the difference}. In other
words, no matter what the preparation of the input systems is and
no matter what observable we measure on the outputs of the
teleportation device, we will always get the same probability
distribution of results as if the inputs were directly measured.
Note also that this criterion does not involve the states of
individual systems, but only states as the distribution parameters
of ensembles of identically prepared systems.

The impossibility of classical teleportation will be treated
extensively in the following section, where it is related to a
hierarchy of impossible machines. For a mathematical statement of
this impossibility in the standard quantum formalism of quantum
mechanics, see the remark after equation~(\ref{class.telepo}). For
the moment, however, let us take it for granted, and see what all
this says about the new concept of quantum information.

First of all, we are concerned here with problems of transmission,
not with content or meaning. This is exactly the same as in
classical information theory. There, too, it is often not easy to
avoid confusion with a different concept of ``information'' used
in everyday language, namely the kind available at an information
desk. Information Theory does not care whether a TV channel is
used for ``misinformation'', but can say everything about what it
takes to secure the technical quality of the final images. Hence
the quantitative measures of ``information'' all relate to storage
and transmission capacity, to the possibilities of compression and
error correction and so on. In the same vein, quantum information
theory will not tell us what the meaning of a ``quantum message''
is, and this is probably meaningless anyway, because a ``read''
message is classical almost by definition. But quantum information
theory has precise notions of the resources needed to transmit
such information faithfully.

Secondly, transmission of quantum information is not at all an
exotic concept in the context of modern physics. It can be
paraphrased in various, perhaps more familiar ways, for example as
``transmission of intact quantum states'', as ``coherent
transmission of quantum systems'' or as transmission ``preserving
all interference possibilities'' of the system. Nevertheless the
information metaphor is useful, not only because it suggests new
applications, but also because it leads one to ask new questions,
and leads to quantitative notions where previously there was only
a qualitative understanding. And possibly this is even a way to
see in a sharper light the old conundrums of the foundations of
quantum mechanics.

\chapter{Impossible Machines\label{sec:impies}}

The usefulness of considering \inx{impossible machines} is well-known
from thermodynamics: the second fundamental law of thermodynamics
is often stated as the impossibility of a perpetual motion
machine. The theorem on the impossibility of classical
teleportation is likewise a fundamental law of quantum mechanics,
and a lot can be learned from analyzing it. Typically, the
impossible machines of quantum theory are perfectly possible in
classical physics, so their impossibility does not follow
superficially from their description, but rather carries a
connotation of paradox.

We will discuss a range of impossible tasks consisting of
\begin{itemize} \item Teleportation \item Copying (``\inx{Cloning}'')
\item Joint Measurement \item Bell's Telephone \end{itemize} As we
will see, Teleportation is the most powerful of these, in the
sense that if we had a teleportation device, we could build a
Quantum Copier, from which we could in turn construct Joint
Measurements, and, finally a device known as \inx{Bell's Telephone}, by
which we could set up superluminal communication. Hence, if we
uphold the principle of Causality, which forbids the weakest
machine in this hierarchy, we are certain that teleportation is
likewise impossible. In this section we will follow this line of
reasoning to prove the impossibility of Teleportation. Of course,
there are other, more direct ways of proving it from the structure
of quantum mechanics. However, these usually require more of the
quantum formalism and give less insight into the differences
between classical and quantum information.

\section{The \inx{Quantum Copier}}

This is the machine referred to in the famous paper of Wootters
and Zurek, entitled ``A single quantum cannot be cloned''
\cite{WooZu}. By definition, a copier would be a device taking one
quantum system as input and turning out two systems of the same
type. The condition for calling this a (faithful) copier is that
we won't be able to distinguish the systems coming from either
output from the input systems by any statistical test, i.e., by
the probabilities measured by any observable, and on any
preparation of initial states. Hence the device has to operate on
arbitrary ``unknown'' states. It is clear that a copier in the
ordinary sense, e.g., a mail relay distributing email to several
recipients, indeed satisfies this condition in the domain of
classical information. Note that we are not so unreasonable as to
demand what the title quoted above suggests, namely that we could
test this device on {\it single events}, or even assume some
ontological ``identity'' of input and output: the criterion for
faithful copying is flatly statistical, and can be verified by a
straightforward collection of statistical tests.

\begin{figure}[htb]
  \begin{center}
    \includegraphics[scale=1]{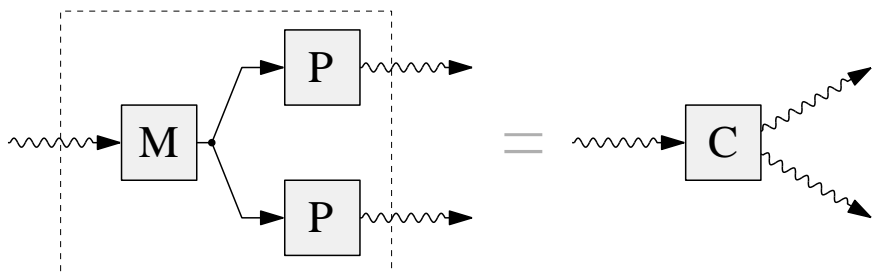}
    \caption{Making a copier from a ``classical teleportation'' line}
    \label{fig:Clone}
  \end{center}
\end{figure}

Given a teleportation device, building a copier is quite easy (see
Figure~\ref{fig:Clone}). All we have to do is to remember that the
classical information obtained in the intermediate stage of the
teleportation process can be copied perfectly. Hence we can apply
the measuring device of the teleportation line to the input
systems, copy the results, and simply run the reconstructing
preparation on each of these copies.

\section{The Joint Measurement}

This is the task of combining two separate measuring devices into
a single device, or the ``simultaneous measurement'' of two
quantum observables $A$ and $B$. Thus a joint measuring device
``$A\&B$'' is a device giving a pair $(a,b)$ of classical outputs
each time it is operated, such that $a$ is a possible output of
$A$, and $b$ is a possible output of $B$. We require that the
statistics of the $a$ outcomes alone is the same as for device
$A$, and similarly for $B$. Note that once again our criterion is
statistical, and can be tested without recourse to \inx{counterfactual
conditionals} such as ``the result which would have resulted if $B$
rather than $A$ had been measured on this particular quantum
particle''.

Many quantum observables are not jointly measurable in this sense.
The most famous examples, position and momentum, different
components of angular momentum, and positions of a free particle
at different times, are probably contained in every quantum
mechanics course. Hence the impossibility of joint measurements is
nothing but a precise statement of an aspect of
``\inx{complementarity}''.

\begin{figure}[htbp]
  \begin{center}
    \includegraphics[scale=1]{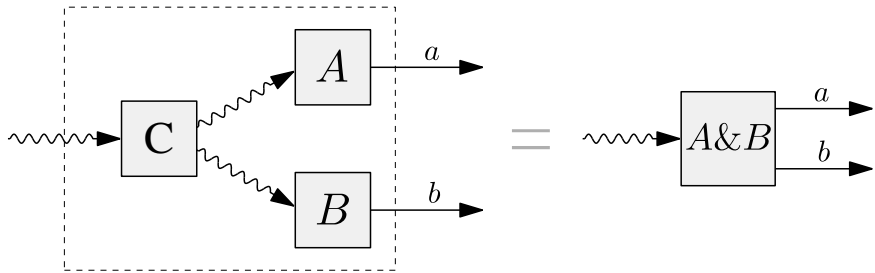}
    \caption{Getting joint measurements from a copier}
    \label{fig:JointM}
  \end{center}
\end{figure}

Nevertheless, a joint measurement device for any of these could
readily constructed given a functioning quantum copier (see
Figure~\ref{fig:JointM}): one would simply run the copier C on the
quantum system, and then apply the two given measuring devices, A
and B,  to the copies. It is easy to see that the definition of
the copier then guarantees that the statistics of $a$ and $b$
separately come out right. In other words, a copier can be seen as
a \itx{universal joint measuring device}.

\section{\inx{Bell's Telephone}\label{sec:belltel}}

This is not named after a certain phone company, but after John S.
Bell, who never proposed it in this form, but might have. It
refers to the project of installing superluminal communication
using only correlations of the type tested by Bell's inequalities.
Without going into details for the moment, the basic setup would
consist of a source producing pairs of particles, and sending one
member of the pair to each of the two communicating parties,
conventionally named ``Alice'' and ``Bob''. Each of them has a
collection of different measuring devices to choose from, and the
idea is for Alice to do something which creates a noticeable
change in the probabilities measured by Bob. Clearly, this is a
paradoxical task, because no particle or other physical carrier of
information actually goes from Alice to Bob. Therefore, if only
the particles move sufficiently far apart, this device would
transmit superluminally.

It is maybe useful to point out here a common confusion concerning
such superluminal effects, which sometimes even afflicts otherwise
reliable professional writers. The mistake is usually spotted
easily by a device I call the ``\itx{Ping Pong Ball Test}''. It
goes like this:
\begin{quote} {\it Take an author's explanation of
Bell's inequalities, and substitute ``ping pong balls'' for every
quantum particle. Then if whatever the author is selling as
paradoxical, remains true, he/she hasn't understood a thing.}
\end{quote}
Here is an example: imagine a box containing a ping pong ball,
which can be separated into two parts, without looking at the
ball. One part is shipped to Tokyo or Alpha Centauri, without
looking inside. Then if I open the other box I know instantly,
i.e., ``at superluminal speed'' whether the ball is at the distant
location or not. Of course, that is true, but hardly paradoxical,
and totally useless for sending a message either way. To repeat:
there is nothing paradoxical in statistical correlations per se
between distant systems with a common past, even if the
correlation is perfect.

If Alice wants to send a message to Bob, correlations between any
two measuring devices are useless, because they cannot even be
detected without comparing the results, which requires exactly the
communication the Telephone was intended for. Only if something
Alice {\it does} has an effect on measuring results at Bob's end
we can speak of communication. The only thing Alice can do in the
standard setup is to choose a measuring device, and Bell's
Telephone can be said to work if these choices have an influence
on the probabilities measured by Bob (who has no access to Alice's
measuring results). If there is no physical system traveling from
Alice to Bob, however, this will be impossible.

To be sure, this can hardly be counted as an impossible machine of
quantum mechanics, since the argument has nothing to do with
quantum theory. What makes it fit into the hierarchy described
here is the following: if we assume that Bob has a joint measuring
device for two yes/no measurements, and \itx{Bell's inequality} is
violated, we can design a strategy for Alice to send signals to
Bob with better than chance results. Hence the joint measurement
of suitable observables  can be a device sufficiently strong to
achieve a task forbidden by Causality, and is hence impossible in
general. This is the last construction  in the hierarchy of
impossible machines mentioned at the beginning of this section.

The proof of this step  amounts to yet another derivation of
Bell's inequalities, but since it emphasizes the communication
aspect it fits well into our context, and we will at least sketch
it. This step will be rather more technical than the rest of this
section, but does not require any quantum theory. The argument can
be skipped without loss to later sections.
\begin{figure}[htbp]
  \begin{center}
    \includegraphics[scale=1]{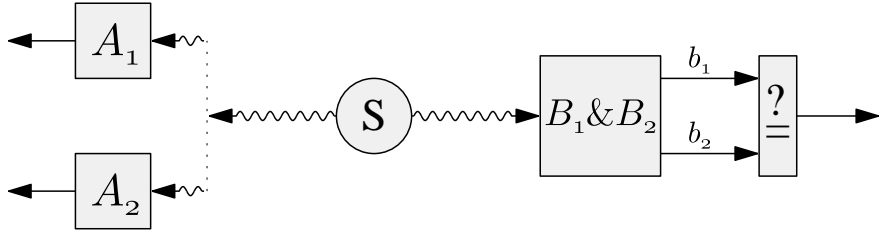}
    \caption{Building Bell's Telephone from a joint measurement}
    \label{fig:BPhone}
  \end{center}
\end{figure}

So let us assume that Alice and Bob each have at their disposal
two measuring devices, say $A_1,A_2$ and $B_1,B_2$, respectively.
Each of these can either give the result $+1$ or $-1$. We will
denote by $\pp a,b/A_i,B_j/$ the probability for Alice to get $a$,
and Bob to get $b$, in a correlation experiment in which Alice
used measuring device $A_i$ and Bob uses $B_j$. By
\begin{displaymath}
    C(A_i,B_j)=\sum_{a,b}ab\ \pp a,b/A_i,B_j/
\end{displaymath}
we will denote the correlation coefficient, which lies between
$-1$ and $+1$. The combination
\begin{equation}\label{e.bellcorr}
  \beta=C(A_1,B_1)+C(A_1,B_2)+C(A_2,B_1)-C(A_2,B_2)
\end{equation}
carries special significance, as we will see below. Because the
inequality ``$\beta\leq2$'' is known as the Bell inequality, we
will call $\beta$ the \itx{Bell correlation} for this choice of
four observables. It is a quantity directly accessible to
experiment. Note that usually Bob cannot tell from his data which
apparatus ($A_1$ or $A_2$) Alice chose. This is reflected by the
equation
\begin{displaymath}
 \sum_a\pp a,b/A_1,B_j/=\sum_a\pp a,b/A_2,B_j/
    \equiv\pp b/B_j/\;,
\end{displaymath}
and borne out by all known experimental data. Now suppose Bob has
a joint measuring device for his $B_1$ and $B_2$, which we will
denote by $B_1\&B_2$, which produces pair outcomes $(b_1,b_2)$
(see Figure~\ref{fig:BPhone}). We can then determine the
probabilities $p_i(a_i,b_1,b_2)=\pp a_i,(b_1,b_2)/A_i,B_1\&B_2/$.
The condition that this is really a joint measurement is expressed
by the equations
\begin{eqnarray}\label{bell-joint}
  \sum_{b_1}p_i(a_i,b_1,b_2)&=&\pp a_i,b_2/A_i,B_2/ \quad\hbox{and }\\
   \sum_{b_2}p_i(a_i,b_1,b_2)&=&\pp a_i,b_1/A_i,B_1/\;,
\end{eqnarray}
each for $i=1,2$. The basic rule for the information transmission
is the following:
\begin{quote}{\sl
Alice encodes the bit she wants to send by either choosing
apparatus $A_1$ or apparatus $A_2$. Then Bob looks at his readout
and interprets it as ``$A_1$'', whenever the two displays coincide
($b_1=b_2$) and as ``$A_2$'', if they are different.}
\end{quote}
We can then estimate the probability $p_{\rm ok}$ for Bob to be
right, assuming that the choices $A_1$ and $A_2$ are made with the
same frequency. Assume first that Alice chooses $A_1$. Then Bob is
right  with probability
\begin{displaymath}
   \sum_{a_1,b_1,b_2}\left\vert\frac{b_1+b_2}2\right\vert\ \abs{a_1}\
        p_1(a_1,b_1,b_2)\;,
\end{displaymath}
where the first factor takes into account the condition $b_1=b_2$,
and the second is introduced for later convenience. Combining this
with the second term of this kind for Alice's choice $A_2$, and
taking into account the probability $1/2$ for these choices we get
the overall probability $p_{\rm ok}$ for Bob to be correct as
\begin{eqnarray}
  p_{\rm ok}&=&\frac12\sum_{a_1,b_1,b_2}
       \left\vert\frac{b_1+b_2}2\right\vert\ \abs{a_1}\ p_1(a_1,b_1,b_2)
  \nonumber\\&&\qquad
           +\frac12\sum_{a_2,b_1,b_2}
          \left\vert\frac{b_1-b_2}2\right\vert\ \abs{a_2}\ p_2(a_2,b_1,b_2)
  \nonumber\\
     &\geq&\frac14\sum_{a_1,b_1,b_2}
      (b_1+b_2)a_1\; p_1(a_1,b_1,b_2)
  \nonumber\\&&\qquad
      +\frac14\sum_{a_2,b_1,b_2}
       (b_1-b_2)a_1\; p_2(a_2,b_1,b_2)
  \nonumber\\
     &=& \frac14\Bigl( C(A_1,B_1)+C(A_1,B_2)+C(A_2,B_1)-C(A_2,B_2)\Bigr)
  \nonumber\\
     &=& \frac\beta4\;.
\end{eqnarray}
Bob is right with better than chance, if $p_{\rm ok}>1/2$, which
by this computation can be guaranteed as soon as $\beta>2$, i.e.,
as soon as the classical Bell inequality (in
Clauser-Horne-Shimony-Holt form \cite{CHSH}) is violated. But this
is indeed the case in the experiments conducted to determine
$\beta$ (e.g., \cite{Aspect}), which give roughly
$\beta\approx2\sqrt2\approx2.8$. If we believe these experiments,
the only conclusion is that the joint measurability of the $B_1$
and $B_2$ used in the experiment would be sufficient to make
Bell's Telephone work, which was our claim.

\section{Entanglement, mixed state analyzers, and correlation resolvers}

Violations of Bell's inequalities can also be seen to prove the
existence of a new class of correlations between quantum systems,
known as \itx{entanglement}. This concept is as fundamental to the
field of quantum information theory as the idea of quantum
information itself. So rather than organizing this introduction as
an answer to the the question ``why quantum information is
different from classical information'', we could have followed the
line ``why entanglement is different from classical correlation''.
There are impossible machines in this line of approach, too, and
we will now describe briefly how they fit in.

Consider a correlation experiment of the kind used in Bell's
inequalities (see Section~\ref{sec:belltel}). If Bob looks at his
particles, and makes measurements on them without any
communication from Alice, he will find that their statistics are
described by a certain mixed state. It must be mixed, because if
he now listens to Alice and sorts his particles according to
Alice's measuring results, he will get two subensembles, which are
in general different. In the usual ideal 2-qubit situation, in
which one gets the maximal violation of Bell's inequalities, these
subensembles are described by pure states.

This is very satisfying for people who see the occurrence of mixed
states in quantum mechanics merely as a result of ignorance, as
opposed to the deeper kind of randomness encoded in pure states.
This view usually comes with an \itx{individual state}
interpretation of quantum mechanics, by which each individual
system can be assigned a pure state (a single vector in Hilbert
space), and a general preparing procedure is not just given by its
density matrix, but by a specific probability distribution of pure
states. Let us call a \itx{mixed state analyzer} a hypothetical
device, which can see the difference, i.e., a measuring device
whose output after many measurements on a given ensemble is not
just a collection of expectations of quantum observables, but the
distribution of pure states in the ensemble. In the case of a
correlation experiment, where Bob sees a mixed state only because
he is ignorant about Alice's results, this machine would find for
him the decomposition of his mixed state into two pure states.

The problem is, of course, that Alice has several choices of
measuring devices, and that the decomposition of Bob's mixed state
depends, accordingly, on Alice's choice. Hence she could signal to
Bob, and we would have another instance of Bell's Telephone. There
would be a way out if Joint Measurements were available (to Alice
in this case): then we could say that the two decompositions were
just the first step in an even finer decomposition, a further
reduction of ignorance, which would be brought to light if Alice
would apply her joint measurement. Presumably the mixed state
analyzer would then yield this finer decomposition, because the
operation of this device would not depend on how closely Alice
cares to look at her particles.

But just as two quantum observables are often not jointly
measurable, two decompositions of mixed states often have no
common refinement (Actually, in the formalism of quantum theory
these are two variants of the same theorem). In particular, the
two decompositions belonging to Alice's choices in an experiment
demonstrating a violation of Bell's inequalities have no common
refinement, and any mixed state analyzer could be used for
superluminal communication in this situation.

Another device, which is suggested by the individual state
interpretation arises from a naive extrapolation of this view to
the parts of a composite system: if every single system can be
assigned a pure state, a composite system could be assigned a pair
of pure states, one for each subsystem. A correlated state should
therefore be given by a probability distribution of such pairs. A
device, which represents an arbitrary state of a composite system
as a mixture of uncorrelated pure product states might be called a
\itx{correlation resolver}. It could be built given a classical
teleportation line: when one applies the teleportation to one of
the subsystems, and conditions on the classical measurement
results of the intermediate stage, one gets precisely a
representation of an arbitrary state in this form. But it is easy
to see that any state which can be so analyzed automatically
satisfies all Bell-type inequalities, and hence once again the
experimental violations of Bell's inequalities show that such a
correlation resolver cannot exist. Hence we have here a second
line of reasoning for showing the No-Teleportation Theorem: a
teleportation device would allow classical correlation resolution,
which is shown to be impossible by the Bell experiments.

The distinction of resolvable states and their complement is one
of the starting points of entanglement theory, where the
``resolvable'' states are called ``separable'', or ``classically
correlated'', and all others or simply ``entangled''. For more
detailed treatment and an up-to date overview, the reader is
referred to the article by the Horodecki family in this volume.

Without going into philosophical discussions on the foundations of
quantum mechanics, I should comment briefly on the individual
state interpretation, which has suggested the two impossible
machines discussed in this subsection. First, this view is not at
all uncommon, and it is quite possible to read some passages from
the Masters of the Copenhagen Interpretation as an endorsement of
this view. Secondly, if we define a \itx{hidden variable theory}
as a theory in which individual systems are described by classical
parameters, whose distribution is responsible for the randomness
seen in quantum experiments, we have no choice but to call the
individual state interpretation a hidden variable theory. The
hidden variable in this theory is usually denoted by $\psi$. And
sure enough, as we have just pointed out, it has all the
difficulties with locality such a theory is known to have on
general grounds. Thirdly, avoiding an individual state
interpretation, and with it some of its misleading intuitions, is
easy enough. In practice this is done anyhow, by concentrating on
those aspects of the theory, which have some direct statistical
meaning, not involving hypothetical, and usually impossible
devices. This common ground is the statistical interpretation of
quantum mechanics, in which states (pure or mixed) are the analogs
of classical probability distributions, and are not seen as a
property of the individual system, but of a specific way of
preparing the systems.

\chapter{Possible Machines}

\section{Operations on multiple inputs}

The No-Teleportation Theorem derived in the previous chapter says
that there is no way to measure a \inx{quantum state} in such a way that
the measuring results suffice to reconstruct the state. At first
sight this seems to deny that the notion of ``quantum states'' has
an operational meaning at all. But there is no contradiction, and
we will resolve the apparent conflict in this subsection, if only
to sharpen the statement of the No-Teleportation Theorem.

Let us recall the operational definition of quantum states,
according to the statistical interpretation of quantum mechanics.
A state is the description of a way of preparing quantum systems,
in all aspects relevant to computing expectation values. We might
also say that it {\it is} the assignment of an expectation value
to every observable of the system. So to the extent that
expectation values can be measured, it is possible to determine
the state by testing it on sufficiently many observables. What is
crucial, however, is that even the determination of a single
expectation value is a {\it statistical} measurement. Hence it
requires a repetition of the experiment many times, using many
systems prepared according to the same procedure. In contrast, the
above description of teleportation demands that it works with a
single quantum system as input, and that the measuring device does
not accumulate results from several input systems. Expressed in
the current jargon: teleportation is required to be a
\itx{one-shot operation}. Note that this does not contradict our
statistical criteria for success of teleportation and other
devices, which involve a statistics of independent ``single
shots''.

\begin{figure}[htb]
  \begin{center}
    \includegraphics[scale=1]{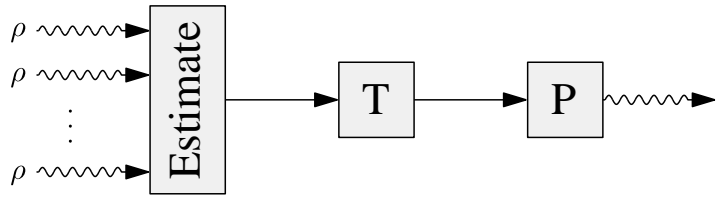}
    \caption{Classical Teleportation on multiple inputs,
       or state estimation}
    \label{fig:estcl}
  \end{center}
\end{figure}

If we have available many identically prepared systems, many
operations which are otherwise impossible, become easy. Let us
begin with classical teleportation. Its multi-input analog is the
\itx{state estimation} problem: how can we design a measurement
operating on samples of many (say, $N$) systems from the same
preparing device, such that the measuring result in each case is a
collection of classical parameters forming a hermitian matrix,
which on average is close to the density matrix describing the
initial preparation. This is symbolized in Figure~\ref{fig:estcl}
(with the box T omitted for the moment): the box P at the end
would be a repreparation of systems according to the estimated
density matrix. The overall output will then be a quantum system,
which can be directly compared with the inputs in statistical
experiments. It is clear that the state cannot be determined
exactly from a sample with finite $N$, but the determination
becomes arbitrarily good in the limit $N\to\infty$. Optimal
estimation observables are known in the case when the inputs are
guaranteed to be pure \cite{estpure}, but in the case of general
mixed states there are no clear cut theorems yet, partly due to
the fact that it is less clear what ``figure of merit'' best
describes the quality of such an estimator.

Given a good estimator we can, of course, proceed to good cloning
by just repeating the re-preparation P as often as desired. The
surprise here \cite{clone1} is that if only a fixed number $M$ of
outputs is required, it is possible to get better clones by
devices staying entirely in the quantum world than by going via
classical estimation. Again, the problem of \itx{optimal cloning}
is fully understood for pure states \cite{clone2}, but work has
only just begun to understand the mixed state case.

Another operation, which becomes accessible in this way is the
\itx{Universal Not} operation, assigning to each pure qubit state
the unique pure state orthogonal to it. Like time reversal, this
is just a special case an anti-unitarily implemented symmetry
operation. In this case, the strategy using a classical estimation
as an intermediate step can be shown to be optimal \cite{unot}. In
this sense ``Universal Not'' is a harder task than ``cloning''.

More generally, we can look at schemes as in
Figure~\ref{fig:estcl}, with T representing any transformation of
the density matrix data, whether or not this transformation
corresponds to a physically realizable transformation of quantum
states. A further interesting application is to the
\itx{purification} of states. In this problem it is assumed that
the input states were once pure, but later corrupted in some noisy
environment (the same for all inputs). The task is to reconstruct
the original pure states. Usually, the the noise corresponds to an
invertible linear transformation on the density matrices, but its
inverse is not a possible operation, because it takes some density
matrices to operators with negative eigenvalues. So the reversal
of noise is not possible by a one-shot device, but is easy to a
high accuracy when many equally prepared inputs are available. In
the simplest case of a so-called depolarizing channel this problem
is well understood \cite{purify1}, also in the version requiring
many outputs as in the optimal cloning problem \cite{purify2}.

\section{Quantum Cryptography}
It may seem impossible to find applications of impossible
machines. But that is not quite true: sometimes the impossibility
of a certain task is precisely what is called for in an
application. A case in point is cryptography: here one tries to
make deciphering of a code impossible. So if we can design a code,
whose breaking would require one of the machines in the previous
section we could guarantee its security {\it with the certainty of
Natural Law}. This is precisely what \itx{Quantum Cryptography}
sets out to do. Because only small quantum systems are involved it
is one of the ``easiest'' applications of quantum information
ideas, and was indeed the first to be realized experimentally. For
a detailed description  we refer to the article by Weinfurter and
Zeilinger. Here we just describe in what sense it is the
application of an impossible machine.

As always in cryptography, the basic situation is that two
parties, Alice and Bob, say, want to communicate without giving an
Evil \inx{Eavesdropper}, conventionally named Eve, a chance to listen
in. What classical eavesdroppers do is to tap the transmission
line, {\it make a copy} of what they hear for later analysis, and
otherwise let the signal pass undisturbed to the legitimate
receiver (Bob). But if the signal is quantum, the \inx{No-Cloning
Theorem} tells us that faithful copying is impossible. So either
Eve's copy or Bob's copy is corrupted. In the first case Eve won't
learn anything, and there was no eavesdropping anyway. In the
second case Bob will know something may have gone wrong, and will
tell Alice that they must discard that part of the secret key they
were exchanging. Of course, intermediate situations are possible,
and one has to show very carefully that there is an exact tradeoff
between the amount of information Eve can get and the amount of
perturbation she must inflict on the channel.

\section{Entanglement assisted Teleportation\label{sec:telepo}}

This is arguably the first major discovery in the field of quantum
information. The No-Cloning and No-Teleportation Theorems,
although not formulated in such terms, would hardly have come as a
surprise to people working on foundations of quantum mechanics in
the sixties, say. But entanglement assistance was really an
unexpected turn. It was first seen by Bennett, Brassard, Crepeau,
Jozsa, Peres, and Wootters \cite{telepo}, who also coined the term
``\inx{teleportation}''. It is gratifying to see, though hardly a
surprise on the same scale, that this prediction of quantum
mechanics has also been implemented experimentally. The
experiments are another interesting story, which will no doubt be
told much better in the article of Weinfurter and Zeilinger, who
represent one team in which major breakthrough in this regard was
achieved.

\begin{figure}[htbp]
  \begin{center}
    \includegraphics[scale=1]{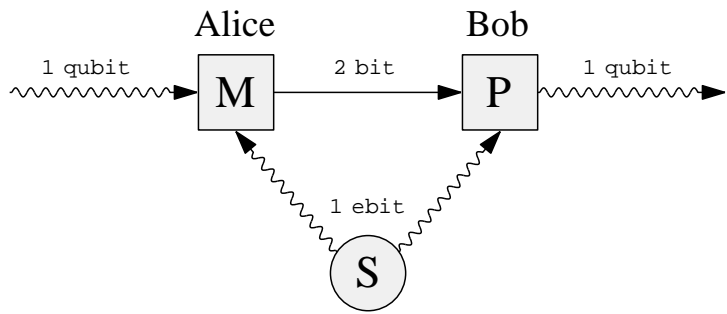}
    \caption{Entanglement assisted Teleportation }
    \label{fig:telepoQ}
  \end{center}
\end{figure}

The teleportation scheme is shown in Figure~\ref{fig:telepoQ}.
What makes it so surprising is that it combines two machines whose
impossibility was discussed in the previous section: omitting the
entanglement distribution (the lower half of
Figure~\ref{fig:telepoQ}) we get the impossible process of
classical teleportation. On the other hand, if we omit the
classical channel, we get an attempt to transmit information on
correlations alone, i.e., a version of Bell's telephone.
Since the time dimension is not represented in this diagram, let
us consider the steps in due order. The first step is that  Alice
and Bob each receive one half of an entangled system. The source
can be a third party or can be Bob's lab. The last choice is maybe
best for illustrative purposes, because it makes clear that no
information is flowing from Alice to Bob at this stage. Alice is
next given the quantum system whose state (unknown to her) she is
to teleport. Alice then makes a measurement on the system combined
out of the input and her half of the entangled system. She sends
the results via a classical channel to Bob, who uses them to
adjust the settings on his device, which then performs some
unitary transformation on his half  of the entangled system. The
resulting system is the output, and if everything is chosen in the
right way, these output systems are indeed statistically
indistinguishable from the outputs. To see just how entangled
state S, measurement M and repreparation P have to be chosen,
requires the mathematical framework of quantum theory. In the
standard example one teleports the state of one qubit, using up
one maximally entangled two qubit system (jargon: ``1 \inx{ebit}'') and
sending two classical bits from Alice to Bob. A general
characterization of the teleportation schemes for qubits and
higher dimensional systems is given below in
Section~\ref{sec:telepo}.

\section{\inx{Superdense Coding}\label{sec:sdcode}}

It is easy to see, and in fact a commonplace occurrence that
classical information can be transmitted on quantum channels. For
example, one bit of classical information can be coded in every 2
level system, like, e.g., the polarization degree of freedom of a
photon. It is not entirely trivial to prove, but hardly surprising
that one cannot do better than ``1 bit per qubit''. Can we beat
this bound using the idea of entanglement assistance? It turns out
that one can. In fact one can double the amount of classical
information carried by a quantum channel (``2 bits per qubit'').
Remarkably, the setups for doing this are closely related to
teleportation schemes, and in the simplest cases Alice and Bob
just have to swap their equipment for entanglement assisted
teleportation. This is explained in detail in
Section~\ref{sec:telepo}.

\section{Quantum Computation}

Again, we will be very brief on this subject, although it is
certainly central to the field. After all, it is partly the
promise of a fantastic new class of computers, which has boosted
the interest in quantum information in recent years. But since in
this book computation is covered in the article by Beth, we will
only make a few remarks, connecting it to the theme of possible
versus impossible machines.

So can Quantum Computers perform otherwise impossible tasks? Not
really, because in principle we can solve the dynamical equations
of quantum mechanics on a classical computer, and simulate all the
results. Hence classical unsolvable problems like the \inx{Halting
Problem} for Turing Machines, or the \inx{Word Problem} in group theory
cannot be solved on quantum computers either. But this argument
only shows the {\it possibility} of emulating all quantum
computations on a classical computer, and omits the fact that the
efficiency of this procedure may be terrible. The great promise of
Quantum Computation lies therefore in the reduction of running
time, in the case of \inx{Shor's} factorization \inx{algorithm} \cite{shor}
from exponential to polynomial time. This reduction is comparable
to replacing the task of  counting all the way up to a 137 digit
number by just having to write it. No matter what the constants
are in the growth laws for the computing time (and they will
probably not be very favorable for the quantum contestant), the
polynomial time is going to win if we are really interested in
factoring very large numbers.

A word of caution is necessary here concerning the
impossible/possible distinction. While it is true that no
polynomial time classical factoring algorithm is known, and this
is what counts from a practical point of view,  there is no {\it
proof} that no such algorithm exists. This is a typical state of
affairs in complexity theory, because the non-existence of an
algorithm is a statement about the rather unwieldy set of all
Turing machine programs. A proof by inspecting all of them is
obviously out, so it would have to be based on some principle of
``conservation of difficulties'', which rarely exists for real
life problems. One problem in which this is possible is
identifying which (unique) element of a large list has a certain
property (``needle in a haystack''). In this case the obvious
strategy of inspecting every element in turn can be shown to be
the optimal classical one, and has a running time proportional to
the length $N$ of the list. But \inx{Grover's} quantum \inx{algorithm}
\cite{grover} does it in the order of $\sqrt N$ steps, an amazing
gain even if it is not exponential. Hence there are problems for
which quantum computers are provably faster than any classical
computer.

So what makes it work? This is not so easy to answer, even after
working through Shor's algorithm and verifying the claim of
exponential speedup. Massive entanglement is used in the
algorithm, so this is certainly one important element. Then there
is a technique known as \itx{quantum parallelism}, in which a
quantum computation is run on a coherent superposition of all
possible classical inputs, and in a sense, all values of a
function are computed simultaneously. A catchy paraphrase
attributed to D.~Deutsch is to call this a computation in the
parallel worlds of the \inx{many-worlds interpretation}.

But perhaps the best way to find out what powers quantum
computation is to to turn it around and to really try the
classical emulation. The practical difficulty which then becomes
apparent immediately  is that Hilbert space dimensions grow
extremely fast. For $N$ qubits (two-level systems) one has to
operate in a Hilbert space of $2^N$ dimensions. The corresponding
space of density matrices has $2^{2N}$ dimensions. For classical
bits one has instead a configuration space of $2^N$ discrete
points, and the analogue of the density matrices, the probability
densities live in a merely $2^N$ dimensional space. Brute force
simulations of the whole system therefore tend to grind to a halt
already on fairly small systems. Feynman was the first to turn
this around: maybe only a quantum system can be used to simulate a
quantum system, and maybe, while we are at it, we can go beyond
simulation and do some interesting computations as well. So
putting it positively: in a quantum system we have exponentially
more dimensions to work with: {\it there is lots of room in
Hilbert space}. The added complexity of quantum vs. classical
correlations, i.e., the phenomenon of entanglement, is also a
consequence of this.

But it is not so easy to use those extra dimensions. For example,
for transmission of classical information an $N$-qubit system is
no better than a classical $N$-bit system. Only the entanglement
assistance of superdense coding brings out the additional
dimensions. Similarly, quantum computers do not speed up every
computation, but are only good at specific tasks where the extra
dimensions can be brought into play.

\section{Error correction}

Again we will only make a few remarks related to the
possible/impossible theme, and refer the reader to T.~Beth's
article in this volume for a deeper discussion. First of all,
error correction is absolutely crucial for the implementation of
quantum computers. Very early in the development the suspicion was
raised that exponential speedup was only possible, if all
component parts of the computer were realized with exponentially
high (hence practically unattainable) precision.

In a classical computer the solution to this problem is
\itx{digitization}: every bit is realized by a bistable circuit,
and any deviation from the two wanted states is restored by the
circuit at the expense of some energy and with some heat
generation. This works separately for every bit, so in a sense
every bit has its own heat bath. But this strategy will not work
for quantum computers: to begin with there is now a continuum of
pure states which would have to be stabilized for every qubit, and
secondly, one heat bath per qubit would quickly destroy
entanglement, and hence make the quantum computation impossible.
There are many indications that entanglement is indeed more easily
destroyed by thermal noise and other sources of errors, summarily
referred to as \itx{decoherence}. For example, a \inx{Gaussian channel}
(this is a special type of infinite dimensional channel) has
infinite capacity for classical information, no matter how much
noise we add. But its quantum capacity drops to zero, if we add
more classical noise than specified by the Heisenberg uncertainty
relations \cite{HolWer}.

A standard technique for stabilizing classical information is
\itx{redundancy}: just send a classical bit three times, and
decide at the end by majority vote which bit to take. It is easy
to see that this reduces the probability of error from order
$\varepsilon$ to order $\varepsilon^2$. But quantum mechanically
this procedure is forbidden by the \itx{No-Cloning Theorem}: We
simply cannot make three copies to start the process.

Fortunately \inx{quantum error correction} is possible in spite of all
these doubts \cite{errcor1}. It also works by distributing the
quantum information over several parallel channels, but does this
in a much more subtle way than copying. Using five parallel
channels one can get a similar reduction of errors from order
$\varepsilon$ to order $\varepsilon^2$ \cite{errcor2}. Much more
has been done, but many open problems remain, for which I refer
once again to the article by Beth.

\chapter{A Preview of the Quantum Theory of Information}

Before we go on in the next section to turn some of the heuristic
descriptions of the previous sections into rigorous mathematical
statements, I will try to give a flavor of the theory to be
constructed, and of its motivations and current state of
development.

Theoretical physics contributes to the field of Quantum
Information Processing in two distinct though interrelated ways.
On the one hand, it is necessary to build {\it theoretical models}
of the systems which are being set up experimentally as candidates
for quantum devices. Of course, any such system will have very
many degrees of freedom, of which only very few are singled out as
the ``qubits'' on which the quantum computation is performed.
Hence it is necessary to analyze to what degree and on what time
scales it is justified to treat the qubit degrees of freedom
separately, and with what errors the desired quantum operation can
be realized in the given system. These questions are crucial for
the realizations of all quantum devices, and require specialized
in-depth knowledge of the appropriate theory, e.g., quantum
optics, solid state theory, or quantum chemistry (in the case of
NMR quantum computing). However, these problems are not what we
want to look at in this article.

We are concerned here with another kind of theoretical work, which
could be called the \itx{Abstract Quantum Theory of Information}.
Recall the arguments in Section~2, where the possibility of
translating between different carriers of (classical) information
was taken as the justification for looking at an abstracted
version, the classical Theory of Information, as founded by
Shannon. While it is true that quantum information cannot be
translated into this framework, and is hence a new kind of
information, translation is often possible (at least in principle)
between different carriers of quantum information. Therefore, we
can make a similar abstraction in the quantum case. To this
abstracted theory all qubits are the same, whether they are
realized as polarization of photons, nuclear spins, excited states
of ions in a trap, modes of a cavity electromagnetic field, or
whatever other realization may be feasible. A large amount of work
is currently devoted to this abstract branch of quantum
information theory, so I will list some of the reasons for this
effort.

\begin{itemize}
\item Abstract quantum theoretical reasoning is how it all
started. In the early papers of Feynman and Deutsch, and the
papers by Bennett and co-workers, it is the structure of quantum
theory itself, which opens up all those new possibilities. No hint
from experiment and no particular theoretical difficulty in the
description of concrete systems prompted this development. Since
the technical realizations are lagging behind so much, the field
will probably remain ``{\it theory driven}'' for some time to
come.

\item If we want to transfer ideas from the Classical Theory of
Information to the Quantum Theory, we will always get abstract
statements. This works quite well for importing good questions.
Unfortunately, however, the answers are most of the time not
transferred so easily.

\item The reason for this difficulty with importing classical
results is that some of the standard probabilistic techniques,
such as conditioning, do not work in quantum theory, or work only
sporadically. This is the same problem that the \inx{Statistical
Mechanics} of quantum many-particle systems is facing in comparison
to its classical sister. The cure can only be the development of
new, genuinely quantum techniques. Preferably these should work in
the widest (hence most abstract) possible setting.

\item One of the fascinating aspects of quantum information is
that features of quantum mechanics, which were formerly seen only
as paradoxical or counter-intuitive are now turned into an asset:
these are precisely the features one is trying to utilize now. But
this means that naive intuitive reasoning tends to come to wrong
results. Until we know much more about Quantum Information we will
need rigorous guiding from a solid conceptual and mathematical
foundation of the theory.

\item When we take as a selling point for, say Quantum Cryptography, that
secrets are protected ``with the security of Natural Law'', the
argument is only as convincing as the {\it proof} reducing this
claim to first principles. Clearly this requires abstract
reasoning, because it must be independent of the physical
implementation of the device the eavesdropper uses. It must also
be completely rigorous in the mathematical sense.

\item Because it does not care about the physical realization of
its ``qubits'', the Abstract Quantum Theory of Information is
applicable to a wide range of seemingly very different system.
Consider,  for example some abstract quantum gate like the
``controlled not'' (C-NOT). From the abstract theory we can hope
to get relevant quality criteria such as the minimal fidelity with
which this has to be implemented for some algorithm to work. So
systems of quite different type can be checked according to the
same set of criteria, and a direct competition becomes possible
(and interesting) between different branches of experimental
physics.

\end{itemize}

\noindent So what will be the basic concepts and features of the
emerging Quantum Theory of Information? The information
theoretical perspective typically generates questions like
\begin{quote} {\it How can a given task of quantum information processing be
performed optimally with the given resources?}
\end{quote}
We have already seen a few typical tasks of quantum information
processing in the previous section and, of course, there are more.
Typical resources for cryptography, quantum teleportation, and
dense coding are entangled states, quantum channels and classical
channels. In error correction and computing tasks, resources are
the size of quantum memory, and the number of quantum operations.
Hence all these notions take on a quantitative meaning.

For example, in entanglement assisted teleportation the entangled
pairs are used up (one maximally entangled qubit pair is needed
for every qubit teleported). If we try to run this with less than
maximally entangled states, we may still ask, how many pairs from
a given preparation device are needed per qubit to teleport a
message of many qubits, say, with error less than $\varepsilon$.
This quantity is clearly a measure of entanglement. But other
tasks may lead to different quantitative measures of entanglement.
Very often it is possible to find inequalities between different
measures of entanglement, and establishing these is again a task
of quantum information theory.

The direct definition of the entanglement measure based on
teleportation, or the quantum information capacity of a channel,
and many similar quantities require an optimization with respect
to all codings and decodings of asymptotically long quantum
messages, which is extremely hard to evaluate. In the classical
case, however, there is a simple formula for the capacity of a
noisy channel, called Shannon's Coding Theorem, which allows us to
compute the capacity directly from the transition probabilities of
a channel. Finding quantum analogs of the \inx{Coding Theorem} (and
similar formulas for entanglement resources) is still one of the
great challenges in quantum information theory.

\chapter{Elements of Quantum Information Theory}

It is probably too early to write a definitive account of Quantum
Information Theory -- there are simply too many open questions.
But the basic concepts are clear enough, and it will be the task
of the remainder of this article to explain them, and use these
sharp definitions to state some of the interesting open problems
in the field. In the limited space available this cannot be done
in textbook-style, with many examples and full proofs (or even
full references) of all the things used on the way. So I will try
to emphasize the main lines, and to set up the basic definitions
using as few primitive concepts as possible. For example, the
capacities of a channel for either classical or quantum
information will be defined on exactly the same pattern. This will
make it easier to establish the relations between these concepts.

The following pages begin with material which every physicist
knows from quantum mechanics courses, although maybe not in this
form. We need to go over it, though,  in order to establish
notation.

\section{Systems and States}

The systems occurring in the theory can be either quantum or
classical, or can be hybrids composed of a classical and a quantum
part. Therefore, we need a mathematical framework covering all
these cases. A good choice is to characterize each type of systems
by its \itx{algebra of observables}. In this article all
observable algebras will be taken to be {\it finite dimensional}
for simplicity. Extensions to infinite dimension are mostly
straightforward, though, and in fact a strength of the algebraic
approach to quantum theory is that it deals not just with infinite
dimensional algebras, but also with systems of infinitely many
degrees of freedom as in quantum field theory
\cite{WerSum,clifton} and statistical mechanics \cite{BraRo}.

The first main type of systems are purely \itx{classical systems},
whose observable algebra is commutative, and can hence be
considered as a space of complex valued functions on a set $X$.
Our standing finiteness assumption requires that $X$ is a finite
set, and the observable algebra $\A$ will be $\C(X)$, the space of
all functions $f:X\to\Cx$. A single \itx{classical bit}
corresponds to the choice $X=\{0,1\}$. On the other hand, a purely
{\it quantum system} is determined by the choice $\A=\B(\Hh)$, the
algebra of all bounded linear operators on the Hilbert space
$\Hh$. The finiteness assumption requires that $\Hh$ has  finite
dimension $d$, so $\A$ is just the space $\M_d$ of complex
$d\times d$-matrices. A \itx{qubit} is given by $\A=\M_2$.

The basic statistical interpretation of the observable algebra is
the same in the quantum and classical case, and hinges on the cone
of positive elements in the algebra. Here $Y$ is called
\itx{positive} (in symbols $Y\geq0$) if it can be written in the
form $Y=X^*X$. Then $Y\in\M_d$ is positive, exactly if it is given
by a positive semidefinite matrix, and $f\in\C(X)$ is positive iff
$f(x)\geq0$ for all $x$. In any observable algebra $\A$, we will
denote by $\idty\in\A$ the identity element.

A \itx{state} $\Phi$ on $\A$ is a positive normalized linear
functionals on  $\A$. That is, $\Phi:\A\to\Cx$ is linear, with
$\Phi(X^*X)\geq0$ and $\Phi(\idty)=1$. Each state describes a way
of preparing systems in all the details, which are relevant for
subsequent statistical measurements on the systems. The
measurements are described by assigning to each outcome of a
device an \itx{effect} $F\in\A$, i.e., an element with $0\leq
F\leq\idty$. The prediction of the theory for the probability of
that outcome, measured on systems prepared according to the state
$\rho$ is then $\rho(F)$.

For explicit computations we will often need to expand states and
elements of $\A$ in a basis. The standard basis in $\C(X)$
consists of the functions $e_x, x\in X$, such that $e_x(y)=1$ for
$x=y$ and zero otherwise. Similarly, if $\phi_\mu\in\Hh$ is an
orthonormal basis of the Hilbert space of a quantum system, we
denote by $e_{\mu\nu}=\ketbra{e_\mu}{e_\nu}\in\B(\Hh)$ the
corresponding ``matrix units''. Then a state $p$ on the classical
algebra $\C(X)$ is characterized by the numbers $p_x\equiv
p(e_x)$, which form a probability distribution on $X$, i.e.,
$p(x)\geq0$ and $\sum_xp(x)=1$.  Similarly, a quantum state $\rho$
on $\B(\Hh)$ is given by the numbers
$\rho_{\mu\nu}\equiv\rho(e_{\nu\mu})$, which form the so-called
\itx{density matrix}. If we interpret them as the expansion
coefficients of an operator
$\widehat\rho=\sum_{\mu\nu}\rho_{\mu\nu}e_{\mu\nu}$, the
\itx{density operator} of $\rho$, we can also write
$\rho(A)=\tr(\widehat\rho A)$.

\begin{figure}
 \begin{center}
    \includegraphics[scale=1]{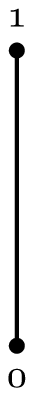}
    \includegraphics[scale=1]{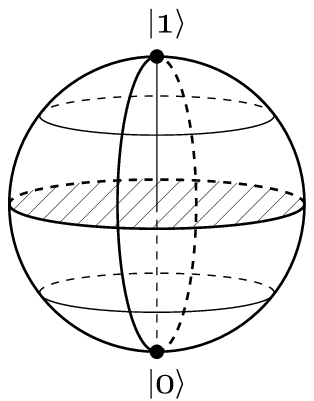}
  \end{center}
\label{fig-sspaces}
    \caption{State spaces as convex sets\hfill\break
    left: one classical bit;
    right: one quantum bit (qubit)}
\end{figure}

A state is called \itx{pure}, if it is extremal in the convex set
of all states, i.e., if it cannot be written as a convex
combination $\lambda\rho'+(1-\lambda)\rho''$ of other states.
These are the states, which contain as little randomness as
possible. In the classical case, the only pure states are those
concentrated on a single point $z\in X$, i.e., $p_z=1$, or
$p(f)=f(z)$. The pure states in the quantum case are determined by
``wave vectors'' $\psi\in\Hh$ such that $\rho(A)=\bra\psi,A\psi>$,
resp. $\widehat\rho=\ketbra\psi\psi$. Thus in the simplest case of
a classical bit there are just two extreme points, whereas in the
case of a qubit the extreme points form a sphere in three
dimensions which are given by the expectations of the three Pauli
matrices: \begin{eqnarray}
  \widehat\rho&=&\frac12\left(
                    \begin{array}{cc}1+x_3&x_1-ix_2\\
                      x_1+ix_2&1-x_3\end{array}\right)
                  =\frac12(\idty+ \vec\sigma\cdot\vec x)\\
            x_k&=& \rho(\sigma_k)\nonumber
\end{eqnarray}
Then positivity requires $\abs{\vec x}^2\leq1$, with equality when
$\rho$ is pure. This is shown in Figure~\ref{fig-sspaces}.

Thus in addition to north pole $\ket+$ and south pole $\ket-$,
which roughly correspond to the extremal states of the classical
bit we have their coherent superpositions corresponding to the
wave vectors $\alpha\ket++\beta\ket-$, with $\alpha,\beta\in\Cx$,
and $\abs\alpha^2+\abs\beta^2=1$. This additional freedom becomes
even more dramatic in higher dimensional systems, and is
crucial for the possibility of entanglement.

Entanglement is a property of states on composite systems, so we
must introduce the notion of \itx{composition} of systems. We will
define this in a way which applies to classical and quantum
systems alike. If $\A$ and $\B$ are the observable algebras of the
subsystems, the  observable algebra of the composition is defined
to be the \inx{tensor product} $\A\otimes\B$. In the finite dimensional
case, which is our main concern, this is defined as the space of
linear combinations of elements written as $A\otimes B$ with
$A\in\A$ and $B\in\B$, such that $A\otimes B$ is linear in $A$ and
linear in $B$. The algebraic operations are defined by $(A\otimes
B)^*=A^*\otimes B^*$, and $(A_1\otimes B_1)(A_2\otimes
B_2)=(A_1A_2)\otimes(B_1B_2)$. Thus
$\idty=\idty_\A\otimes\idty_\B$. Since positivity is defined in
terms of star-operation (adjoint) and product, these definitions
also determine the states and effects of the composite system.

Let us explore how this unifies the more common definitions in the
classical and quantum case. For two classical factors
$\C(X)\otimes\C(Y)$ a basis is formed by the elements $e_x\otimes
e_y$, so the general element is expanded as
\[f=\sum_{x,y}f(x,y)e_x\otimes e_y\;,\]
so that each element can be identified with a function on the
cartesian product $X\times Y$. Hence
$\C(X)\otimes\C(Y)\cong\C(X\times Y)$. Similarly, in the purely
quantum case we can expand in matrix units, and get quantities
with four indices:
$(A\otimes{B})_{\mu\nu,\mu'\nu'}=A_{\mu\mu'}B_{\nu\nu'}$. In a
basis-free way, i.e., when $A,B$ are considered as operators on
Hilbert spaces $\Hh_A,\Hh_B$, this is defined by the equation
\[ (A\otimes B)(\phi\otimes \psi)=(A\phi)\otimes(B\psi)\;,\]
where $\phi\in\Hh_A$ and $\psi\in\Hh_B$, and the tensor product of
Hilbert spaces is formed in the usual way. Hence
$\B(\Hh_A)\otimes\B(\Hh_B)\cong\B(\Hh_A\otimes\Hh_B)$.

But the definition of composition by tensor product of observable
algebras also determines how a quantum-classical \itx{hybrid} must
be described. Such systems occur frequently in Quantum Information
Theory, whenever a combination of classical and quantum
information is given. We will approach hybrids in two equivalent
ways, which are also useful more generally. Suppose we only know
that the first subsystem is classical without assumptions on the
nature of the second, i.e., we want to characterize tensor
products of the form $\C(X)\otimes\B)$. Then every element can be
expanded in the form $B=\sum_xe_x\otimes B_x$, where now
$B_x\in\B$. Clearly, the elements $B_x$ determine $B$, and hence
we can identify the tensor product with the space (sometimes
denoted by $\C(X;\B)$) of $\B$-valued functions on $X$ with
pointwise algebraic operations. Similarly, assume we only know
that $\B=\M_d$ is the algebra of $d\times d$-matrices. Then
expanding in matrix units we find that
$A=\sum_{\mu\nu}A_{\mu\nu}\otimes e_{\mu\nu}$ with
$A_{\mu\nu}\in\A$. That is, we can identify $\A\otimes\M_d$ with
the space (sometimes denoted by $\M_d(\A)$) of $d\times
d$-matrices with entries from $\A$. By using the relation
$e_{\mu\nu}e_{\alpha\beta}=\delta_{\nu\alpha}e_{\mu\beta}$ one
readily verifies that the product in $\A\otimes\M_d$ indeed
corresponds to the usual matrix multiplication in $\M_d(\A)$, with
due care given to the order of factors in products with elements
from $\A$, if $\A$ happens to be non-commutative. The adjoint is
given by $(A^*)_{\mu\nu}=(A_{\nu\mu})^*$. Hence a hybrid algebra
$\C(X)\otimes\M_d$ can be viewed either as the algebra of
$\C(X)$-valued $d\times d$-matrices, or as the space of
$\M_d$-valued functions on $X$.

The physical interpretation of a composite system $\A\otimes\B$ in
terms of states and effects is straightforward. When $F\in\A$ and
$G\in\B$ are effects, so is $F\otimes G$, and this is interpreted
as the joint measurement of $F$ on the first and $G$ on the second
subsystem, where the ``yes'' outcome is taken as ``both effects
give yes''. In particular, $F\otimes\idty_\B$ corresponds to
measuring $F$ on the first system, completely ignoring the second.
Thus, for any state $\rho$ on $\A\otimes\B$ we define the
\itx{restriction} $\rho_\A$ of $\rho$ to $\A$ by
$\rho_\A(A)=\rho(A\otimes\idty_\B)$. In the classical case the
probability density for $\rho_A$ is obtained by integrating out
the $\B$-variables. In the quantum case it corresponds to the
partial trace of density matrices with respect to $\Hh_\B$. In
general, it is not possible to reconstruct the state $\rho$ from
the restrictions $\rho_A$ and $\rho_B$, which is another way of
saying that $\rho$ also describes correlations between the
systems. However, given $\rho_A$ and $\rho_B$, there is always a
state with these restrictions, namely the tensor product
$\rho_A\otimes\rho_B$, which corresponds to an independent
preparation of the subsystems.

A fundamental difference between quantum and classical
correlations lies in the nature of pure states of composite
systems. Classically this is easy: a pure state on the composite
systems $\C(X)\otimes\C(Y)\cong\C(X\times Y)$ is just a point
$(x,y)\in X\times Y$. Obviously, the restrictions of this state
are the pure states concentrated on $x$ and $y$, respectively.
More generally, whenever one of the algebras in $\A\otimes\B$ is
commutative, every pure state will restrict to pure states on the
subsystems.  Not so in the purely quantum case. Here the pure
states are given by unit vectors $\Phi$ in the tensor product
$\Hh_A\otimes\Hh_B$, and unless $\Phi$ happens to be of the
special form $\phi_A\otimes\phi_B$ (and not a linear combination
of such vectors), the state will not be a product, and the
restrictions will not be pure. The following standard form of
vectors in a tensor product, known as the \itx{Schmidt
decomposition}, is used in entanglement theory every day, and
twice on \inx{Sundays}.

\begin{lemma} (1) (``Schmidt Decomposition'')
Let $\Phi\in\Hh_A\otimes\Hh_B$ be a unit vector, and let
$\widehat\rho_A$ denote the density operator of its restriction to
the first factor. Then if $\widehat\rho_A=\sum_\mu\lambda_\mu
\ketbra{e_\mu}{e_\mu}$ (with $\lambda_\mu>0$) is the spectral
resolution, we can find an orthonormal system $e_\mu'\in\Hh_B$
such that
$$ \Phi=\sum_\mu \sqrt{\lambda_\mu}\ e_\mu\otimes e'_\mu\;.$$
(2) (``\inx{Purification}'') An arbitrary quantum state $\rho$ on $\Hh$
can be extended to a pure state on a larger system with Hilbert
space $\Hh\otimes\Hh_B$. Moreover, the restricted density matrix
$\widehat\rho_B$ can be chosen to have no zero eigenvalues, and
with this additional condition the space $\Hh_B$ and the extended
pure state are unique up to a unitary transformation.
\end{lemma}

\begin{proof}
(1) We may expand $\Phi$ as $\Phi=\sum_\mu e_\mu\otimes\psi_\mu$, with
suitable vectors $\Psi_\mu\in\Hh_B$. The reduced density matrix is
determined by
 $$\tr(\widehat\rho_A F)
   =\bra\Phi,(A\otimes\idty)\Phi>
   =\sum_{\mu\nu}\bra e_\mu,Ae_\nu>\bra\psi_\mu,\psi_\nu>
   =\sum_\mu\lambda_\mu\ \bra e_\mu,Ae_\mu>\;.$$
Since $A$ is arbitrary (e.g., $A=\ketbra{e_\alpha}{e_\beta}$),
we may compare coefficients and get
$\bra\psi_\mu,\psi_\nu>=\lambda_\mu\delta_{\mu\nu}$. Hence
$e_\mu'=\lambda^{-1/2}\psi_\mu$ is the desired orthonormal system.

(2) Existence of the purification is evident by defining $\Phi$ as
above, with the orthonormal system $e'_\mu$ chosen in an arbitrary
way. Then $\widehat\rho_B=\sum_\mu\lambda_\mu \ketbra{e'_\mu}{e'_\mu}$, and
the above computation shows that choosing the basis $e_\mu$ is the
only freedom in this construction. But any two bases are linked by
a unitary transformation.
\end{proof}

A non-product pure state is a basic example of an entangled state
in the sense of the following definition:

\begin{definition}\label{d:separable}
A state $\rho$ on $\A\otimes\B$ is called \itx{separable}
(or ``\inx{classically correlated}'') if it can be written as
$$ \rho=\sum_\mu\lambda_\mu\ \rho^A_\mu\otimes\rho^B_\mu\;,$$
with states $\rho^A_\mu$, $\rho^B_\mu$ on $\A$ and $\B$,
respectively, and weights $\lambda_\mu>0$. Otherwise, $\rho$ is
called \itx{entangled}.
\end{definition}

Thus a classically correlated state may well contain non-trivial
correlations. In fact, if either $\A$ or $\B$ is classical, {\it
every} state is classically correlated. What the definition
expresses is only that we may generate these correlations by a
purely classical mechanism: a classical random generator, which
produces the result ``$\mu$'' with probability $\lambda_\mu$,
together with two preparing devices operating independently but
receiving instructions from the random generator: $\rho^A_\mu$ is
the state produced by the $\A$-device if it gets the input
``$\mu$'' from the random generator, and similarly for $\B$. Then
the overall state prepared by this setup is $\rho$, and clearly
the source of all correlations in this state lies in the classical
random generator.

For an extensive treatment of these concepts the reader is now
referred to the contribution by the Horodecki family in this
volume. We will turn instead to the second fundamental type of
objects in quantum information theory, the channels.

\section{Channels}

Any processing step of quantum information is represented by a
``\inx{channel}''. This covers a great variety of operations, from
preparations to time evolutions, measurements, and measurements
with general state changes. Both input or output of a channel may
be an arbitrary combination of classical and quantum information.
The combination of different kinds of inputs or outputs causes no
special problems of formulation: it simply means that the
observable algebras of input and output system of a channel must
be chosen as suitable tensor products.

The basic idea of the mathematical description each channel is to
characterize  $T$ in terms of the way it modifies subsequent
measurements. Suppose the channel converts systems with algebra
$\A$ into systems with algebra $\B$. Then by applying first the
channel, and then a yes/no measurement $F$ on the  $\B$-type
output system, we have effectively measured an effect on the
$\A$-type system, which will be denoted by $T(F)$. Hence a channel
is completely specified by a map $T:\B\to\A$, and we will say, for
simplicity, that this map {\it is} the channel. There is, of
course an alternative way of viewing a channel, namely as a map
taking input states to output states, i.e., states on $\A$ into
states on $\B$, which we we will denote by $T_*$. We will say that
$T$ describes the channel in the \itx{Heisenberg picture}, whereas
$T_*$ describes the same channel in the \itx{Schr\"odinger
picture}. They are connected by the equation
\begin{equation}\label{T*-T}
  \left(T_*(\rho)\right)(F)=\rho(T(F))\;
\end{equation}
where $\rho$ is an arbitrary state on $\A$, and $F\in\B$ is also
arbitrary. The notation on the left hand side is sometimes a bit
clumsy, therefore we will often write $T_*(\rho)=\rho\circ T$,
where ``$\circ$'' denotes composition of maps, in this case from
$\B$ to $\A$ to $\Cx$. A \itx{composition of channels} will then
also be written as $S\circ T$. This has the advantage that things
are written from left to right in the order in which they happen:
first the preparation then some channels, and finally the yes/no
measurement $F$. As a further simplification, we will often follow
the convention of dropping the parentheses of the arguments of
linear operators (e.g., $T(A)\equiv TA$) and dropping the
$\circ$-symbols, but re-introducing any of these elements for
punctuation whenever they help to make expressions unambiguous, or
just more readable.

For many questions in Quantum Information Theory it is crucial to
have a precise notion of the set of possible channels between two
types of systems: clearly, the distinction between ``possible''
and ``\inx{impossible'' machines} in Section~\ref{sec:impies} is of this
kind, but also the search for the ``optimal device'' performing a
certain task. There are two different approaches for defining the
set of maps $T:\B\to\A$ which should qualify as channels, and
luckily they agree. The first approach is {\it axiomatic}: one
just lists the properties of $T$ which are forced on us by the
statistical interpretation of the theory. The second approach is
{\it constructive}: one lists operations which can actually be
performed according to the conventional wisdom of quantum
mechanics and defines the admissible channels as those, which can
be assembled from these building blocks. The equivalence between
these approaches is one of the fundamental Theorems in this field,
known as the Stinespring Dilation Theorem. We will state it after
describing both approaches, and giving a formal definition of
``channels''.

Note that the left hand side of (\ref{T*-T}) is linear in $F$,
which reflects the fact that a mixture of effects (``use effect
$F_1$ in 42\% of the cases and $F_2$ in the remaining cases'')
directly becomes the \inx{mixture} of the corresponding probabilities.
Therefore, the right hand side also has to be linear in $F$, i.e.,
$T:\B\to\A$ must be a linear operator by the statistical
interpretation of the theory. Obviously, it also has to take
positive operators $F$ into a positive $T(F)$, (``$T$ is
positive'') and the trivial measurement remains trivial:
$T\idty_\B=\idty_\A$ (``$T$ is unit preserving, or  unital'').
This is equivalent to $T_*$ being likewise a positive linear
operator with the normalization condition $\tr T_*(\rho)=\tr\rho$.
Finally, we would like to have an operation of ``running two
channels in parallel'', i.e., we would like to define $T\otimes
S:\A_1\otimes\B_1\to\A_2\otimes\B_2$ for arbitrary channels
$T:\A_1\to\A_2$ and $S:\B_1\to\B_2$. Since the identity $\id_n$ on
an $n$-level quantum system $\M_n$ is one of the channels we want
to consider, we must demand that $T\otimes\id_n$ also takes
positive elements to positive elements. This ``\inx{complete
positivity}'' of $T$ is a non-trivial requirement for maps between
quantum systems. If $\A$ or $\B$ is classical, any positive linear
map from $\A$ to $\B$ is automatically completely positive. For
arbitrary completely positive maps the product $T\otimes S$ is
defined and again completely positive, so just requiring
tensorability with the ``innocent bystander'' $\id_n$ suffices to
make all parallel channels well-defined.

\begin{definition} A {\bf channel} converting
systems with observable algebra $\A$ to systems with observable
algebra $\B$ is a completely positive, unit preserving linear
operator $T:\B\to\A$. \end{definition}

In the ``constructive'' approach one allows only maps, which can
be built from the basic operations of (1) tensoring with a second
system in a specified state, (2) unitary transformation, and (3)
reduction to a subsystem. Let us describe these, and some other
basic channels more formally, if only to show the richness of this
concept. We leave the verification of the channel properties,
including complete positivity, to the reader.

\begin{itemize}
\def\labelitemi{$\bullet$}\itemsep=12pt plus 2pt

\item\label{ex.expand}\itx{Expansion}\\
Expands $\A$-system by a $\B$-system in the state $\rho'$, say.
Thus $T_*(\rho)=\rho\otimes\rho'$, or by (\ref{T*-T}),
$T:\A\otimes\B\to\A$ with $T(A\otimes B)=\rho'(B)A$.

\item\label{ex.restrict}\itx{Restriction}\\
In the Heisenberg picture the operation of discarding system $\B$
from the composite system $\A\otimes\B$ is $T:\A\to\A\otimes\B$,
with $T(A)=A\otimes\idty_\B$. As noted before, this corresponds to
taking partial traces if $\B$ is quantum, and to an integration
over $Y$, if $\B=\C(Y)$ is classical.

\item\label{ex.automo}\itx{Symmetry}\\ By definition, the
symmetries of a quantum system with observable algebra $\A$ are
the invertible channels, i.e., channels $T:\A\to\A$ such that
there is a channel $S$ with $ST=TS=\id_\A$. It turns out that
these are precisely the automorphisms of $\A$, i.e., invertible
linear maps $T:\A\to\A$ such that $T(AB)=T(A)T(B)$, and
$T(A^*)=T(A)^*$. For a pure quantum system the symmetries are
precisely the unitarily implemented maps, i.e., the maps of the
form $T(A)=UAU^*$, with $U$ a unitary element of $\A$. To readers
familiar with Wigner's Theorem (e.g., Corollary 3.3. in
\cite{davies}) another class of maps is conspicuously absent here,
namely positive maps of the form $T(A)=\Theta A^*\Theta^*$ with
$\Theta$ \itx{anti-unitary}. It is well known that due to the
positivity of energy a \itx{time-reversal} symmetry  can only be
implemented by such an anti-unitary transformation. But since such
symmetries are not {\it completely} positive, they can only be
global symmetries, and can never occur as symmetires affecting
only a subsystem of the world.

\item\label{ex.measure}\itx{Observable}\\
A measurement is simply a channel with classical output algebra,
say $\B=\C(X)$. Obviously, $T:\B\to\A$ is uniquely determined by
the collection of operators $F_x:=T(e_x)$ via $Tf=\sum_xf(x)F_x$.
The channel property of $T$ is equivalent to
$$ F_x\in\A \;,\quad F_x\geq0  \;,\quad \sum_xF_x=\idty_\A\;.$$
Either the ``resolution of the identity'' $\{F_x\}$ or the channel
$T$ will be called an observable. This differs in two ways from
the usual textbook definitions of this term: firstly, the outputs
$x\in X$ need not be real numbers, and secondly the operators
$F_x$, whose expectations are the probabilities for obtaining
output $x$, need not be projection operators. This is sometimes
expressed by calling $T$ a \inx{generalized observable}, or a
\inx{POVM}, for positive operator valued measure. This is to
distinguish them form the old style ``non-generalized''
observables, which are called \inx{PVM}'s, projection values
measures, because $F_x^2=F_x$.

\item\label{ex.separable}{\it \inx{Separable Channel}}\\
A classical teleportation scheme is the composition of an observable
and a preparation depending on a classical input, i.e., it is of
the form
\begin{equation}\label{class.telepo}
   T(A)=\sum_x F_x\otimes \rho_x(A)   \;,
\end{equation}
where the $F_x$ form an observable, and $\rho_x$ is the
reconstructed state when the measuring result is $x$.
Equivalently, we can say that $T=RS$ where `input of $S$'=`output
of $R$' is a classical system with observable algebra $\C(X)$. The
impossibility of classical teleportation in this language is the
statement that no separable channel can be equal to the identity.

\item\label{ex.instru}\itx{Instrument}\\
An observable describes only the statistics of the measuring
results, but contains no information about the state of the system
after the measurement. If we want such a more detailed
description, we have to count the quantum system after the
measurement as one of the outputs. Thus we get a composite output
algebra $\C(X)\otimes\B$, where $X$ is the set of classical
outcomes of a measurement, and $\B$ describes the output systems,
which can be of a different type in general form the input systems
with observable algebra $\A$. The term ``instrument'' for such
devices was coined by Davies \cite{davies}. As in the case of
observables, it is convenient to expand in the basis $\{e_x\}$ of
the classical algebra. Thus $T:\C(X)\otimes\B\to\A$ can be
considered as a collection of maps $T_x:\B\to\A$, such that
$T(f\otimes B)=\sum_xf(x)T_x(B)$. The conditions for $T_x$ are
$$ T_x:\B\to\A\quad\hbox{\inx{completely positive}, and}\quad
   \sum_xT_x(\idty)=\idty\;.$$
Note that an instrument has two kind of ``marginals'': we can
ignore the $\B$-output, which leads to the observable
$F_x=T_x(\idty_\B)$, or we can ignore the measuring results, which
gives the overall state change $\Bar T=\sum_xT_x:\B\to\A$.

\item\label{ex.vonNeum}\itx{Von Neumann Measurement}\\
A special instrument is a von Neumann measurement, associated with
a family of orthogonal projections, i.e., $p_x\in\A$ with
$p_x^*p_y=\delta_{xy}p_x$, and $\sum_xp_x=\idty$. These define an
instrument $T:\C(X)\otimes\A\to\A$ via $T_x(A)=p_xAp_x$. What von
Neumann actually proposed \cite{vNeu} was the version of this with
one-dimensional projections $p_x$, so the general case is
sometimes called an incomplete von Neumann measurement, or a
L\"uders measurement. The characteristic properties of such
measurements is their \itx{repeatability}: since $T_xT_y=0$ for
$x\neq y$, repeating the measurement a second time (or any number
of times) will always give the same output. For this reason the
``projection postulate'' demanding that any decent measurement
should be of this form dominated the theory of quantum measuring
processes for a long time.

\item\label{ex.prepare}{\it Classical Input}\\
Classical in formation may occur as the input of a device just as
well as in the output. Again this leads to a family of maps
$T_x:\B\to\A$ such that $T:\B\to\C(X)\otimes\A$, with
$T(B)=\sum_xe_x\otimes T_x(B)$. The conditions on $\{T_x\}$ are
$$ T_x:\B\to\A\quad\hbox{completely positive, and}\quad
   T_x(\idty)=\idty\;.$$
Note that this looks very similar to the conditions for
instruments, but the normalization is different. An interesting
special case is a ``preparator'', for which $\A=\Cx$ is trivial.
This prepares $\B$-states depending in an arbitrary way on the
classical input $x$.

\item\label{ex.Kraus}\itx{Kraus Form}\\
Consider quantum systems with Hilbert spaces $\Hh_A$ and $\Hh_B$,
and let $K:\Hh_A\to\Hh_B$ be a bounded operator. Then the map
$T_K(B)=K^*BK$ from $\B(\Hh_B)$ to $\B(\Hh_A)$ is positive.
Moreover, $T_K\otimes\id_n$ can be written in the same form with
$K$ replaced by $K\otimes\idty_n$. Hence $T_K$ is completely
positive. It follows that maps of the form
\begin{equation}\label{e.Kraus}
  T(B)=\sum_xK_x^*BK_x\;  ,\quad\hbox{with\ }
   \sum_xK_x^*K_x=\idty
\end{equation}
are channels. It will be a consequence of the Stinespring Theorem
that {\it any} channel $\B(\Hh_B)$ to $\B(\Hh_A)$ can be written
in this form, which we call the Kraus form following current
usage. This refers to the book \cite{Kraus}, which is a still to
be  recommended early account of the notion of complete positivity
in physics.

\item\label{ex.ancilla}\itx{Ancilla Form}\\
As announced above, every channel, defined abstractly as a
completely positive normalized map can be constructed in terms of
simpler ones. A frequently used decomposition is shown in
Figure~\ref{fig-ancilla}: The input system is coupled to an
auxilliary system A, conventionally called the ``ancilla''
(maid-servant). Then a unitary transformation is carried out,
e.g., by letting the system evolve according to a tailor-made
interaction Hamiltonian, and finally the ancilla
(or, more generally, a suitable subsystem) is discarded.
\end{itemize}

\begin{figure}
 \begin{center}
    \includegraphics[scale=1]{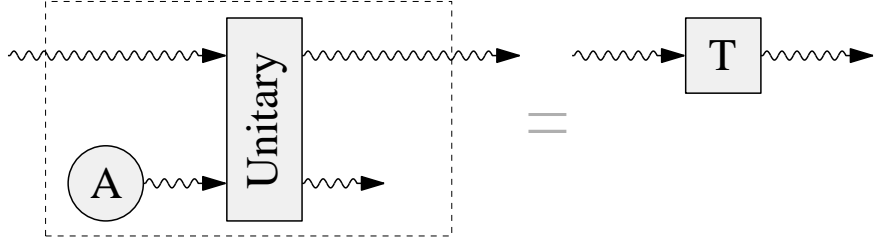}
  \end{center}
\label{fig-ancilla}
    \caption{Representing an arbitrary channel as
    unitary transformation on a system extended by an ancilla.}
\end{figure}

The claim that {\it every} channel can be represented in the last
two forms is a direct consequence of the fundamental structure
theorem for completely positive maps, due to Stinespring
\cite{stine}. We state it here in a version adapted to pure
quantum systems, containing no classical components.

\begin{theorem}\label{Sti} (\inx{Stinespring Theorem}) Let
$T:\M_n\to\M_m$ be a completely positive linear map. Then there is
a number $\ell$, and an operator $V:\Cx^m\to\Cx^n\otimes\Cx^\ell$
such that
\begin{equation}
  T(X)=V^*(X\otimes\idty_\ell)V\;,
\end{equation}
and the vectors of the form
$(X\otimes\idty_\ell)V\phi$ with $X\in\M_n$ and $\phi\in\Cx^m$
span $\Cx^n\otimes\Cx^\ell$. This decomposition is unique up to a
unitary transformation of $\Cx^\ell$. \end{theorem}

The ancilla form of a channel $T$ is obtained by tensoring the
Hilbert spaces  $\Cx^m$ and $\Cx^n\otimes\Cx^\ell$ with suitable
tensor factors $\Cx^a$ and $\Cx^b$, so that $ma=n\ell b$. One
picks pure states in $\psi_a\in\Cx^a$  and $\psi_b\in\Cx^b$ and
looks for a unitary extension of the map $\widetilde
V\phi\otimes\psi_a=(V\phi)\otimes\psi_b$. There are many ways to
do this, and this is a weakness of the ancilla approach in
practical computations: one is always forced to specify an initial
state $\psi_a$ of the ancilla, and many matrix elements of the
unitary interaction, which in the end drop out of all results. As
the uniqueness clause in the Stinespring Theorem shows, it is the
isometry $V$ which neatly captures the relevant part of the
ancilla picture.

In order to get the Kraus form of a general positive map $T$
from its Stinespring representation choose
vectors $\phi_x\in\Cx^\ell$ such that
\begin{equation}\label{Kraus:vec}
  \sum_x\ketbra{\chi_x}{\chi_x}=\idty\;,
\end{equation}
and define Kraus operators $K_x$ for $T$ by
$\bra\phi,K_x\psi>=\bra\phi\otimes \chi_x,V\psi>$
(we leave the straightforward verification of (\ref{e.Kraus})
to the reader). Of course,
we can take the $\chi_x$ as an orthonormal basis of $\Cx^\ell$,
but overcomplete systems of vectors do just as well.

It turns out that {\it all} Kraus decompositions of a given
completely positive operator are obtained in the way just
described. This follows from the following theorem, which solves
the more general problem of finding all decompositions of a given
completely positive operator into completely positive summands. In
terms of channels this problem has the following interpretation:
For an instrument $\{T_x\}$ the sum $\Bar T=\sum_xT_x$ describes
the overall state change, when the measuring results are ignored.
So the reverse question is to find all measurements which are
consistent with a given overall state change (perturbation) of the
system, or in physical terms all \itx{delayed choice measurements}
consistent with a given interaction between system and
environment. By analogy with results for states on abelian
algebras (probability measures) and states on C*-algebras we call
it a Radon-Nikodym Theorem. For a proof see \cite{arveson}.

\begin{theorem}\label{RaNo}
(\inx{Radon-Nikodym Theorem}) Let $T_x:\M_n\to\M_m$, $x\in X$ be a
family of completely positive maps, and let
$V:\Cx^m\to\Cx^n\otimes\Cx^\ell$ be the Stinespring operator of
$\Bar T=\sum_xT_x$.\hfill\break%
Then there are uniquely determined positive operators
$F_x\in\M_\ell$ with $\sum_xF_x=\idty$ such that
$$   T_x(X)=V^*(X\otimes F_x)V
\quad.$$
\end{theorem}

A simple but important special case is the case $\ell=1$: Then
since $\Cx^n\otimes\Cx\equiv\Cx^n$ we can just omit the tensor
factor $\Cx^\ell$. The Stinespring form is then exactly that of a
single term in the Kraus form with Kraus operator $K=V$. The Radon
Nikodym part of the Theorem then says that the only decompositions
of $\Bar T$ into completely positive summands are decompositions
into positive multiples of $\Bar T$. Such maps are also called
``pure''. Since the identity, and more generally symmetries are of
this type we get the following Corollary:

\begin{corollary}\label{niwp}(``No information without \inx{perturbation}'')\\
Let $T:\C(X)\otimes\M_n\to\M_n$ be an instrument with unitary
global state change $\Bar T(A)=T(1\otimes A)=U^*AU$. Then there is
a probability distribution $p_x$  such that $T_x=p_x\Bar T$,  and
the probability $\rho(T_x(\idty))\equiv p_x$ for obtaining
measuring result $x$ is independent of the input state $\rho$.
\end{corollary}

\section{Duality between Channels and Bipartite States}

There are many connections between the properties of states on
bipartite systems and channels. For example, if Alice has locally
created a state, and wants to send one half to Bob, the property
of the channel available for that transmission are crucial for the
kind of distributed entangled state they can be create in this
way. For example, if the channel is separable, the state will also
be separable.

\begin{figure}[htbp]
  \begin{center}
    \includegraphics[scale=1]{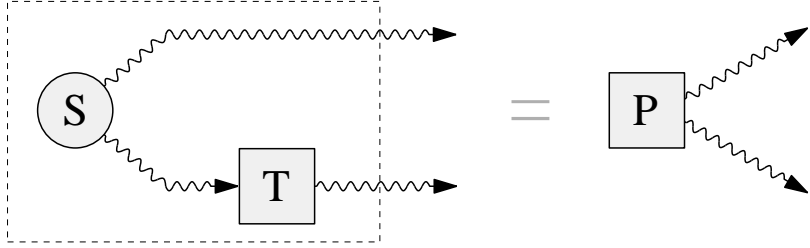}
    \caption{The duality scheme of Lemma~\ref{l.dual}:
       an arbitrary preparation P is uniquely represented
       as preparation S of a pure state and the application
       of a channel $T$ to half of the system. }
    \label{fig:dual}
  \end{center}
\end{figure}

Mathematically, the kind of relationship we will describe here is
very reminiscent of the relationship between bilinear forms and
linear operators: an operator from an $n$-dimensional vector space
to an $m$-dimensional vector space is parametrized by an $n\times
m$-matrix, just like a bilinear form with arguments from an
$n$-dimensional and an $m$-dimensional space. It is therefore
hardly surprising that the matrix elements of density operator on
a tensor product can be reorganized and reinterpreted as the
matrix elements of an operator between operator spaces. What is
perhaps not so obvious, however, is that the positivity conditions
for states and for channels exactly match up in this
correspondence. This is the content of the following Lemma,
graphically represented in Figure~\ref{fig:dual}.

\begin{Lem}\label{l.dual}
Let $\rho$ be a density operator on $\Hh\otimes\K$.
Then there is a Hilbert space $\Hh'$, a pure state $\sigma$ on
$\Hh\otimes\Hh'$, and a channel $T:\B(\K)\to\B(\Hh')$ such that
\begin{equation}\label{e:dual}
  \rho=\sigma\circ(\id_\Hh\otimes T)\;.
\end{equation}
Moreover, the restriction of $\sigma$ to $\Hh'$ can be chosen to
be non-singular, and in this case the decomposition is unique in
the sense that any other decomposition
$\rho=\sigma'\circ(\id_\Hh\otimes T')$ is of the form
$\sigma'=\sigma\circ R$ and $T'=R^{-1}T$ with a unitarily
implemented channel $R$.
\end{Lem}

It is clear that $\sigma$ must be the \inx{purification} of $\rho$,
restricted to the first factor. Thus we may set
$\sigma=\ketbra\Psi\Psi$, with
$\Psi=\sum_k\sqrt{r_k}\; e_k\otimes e'_k$,
where $r_k>0$ are the non-zero eigenvalues of the restriction
of $\rho$ to the first system, and $e'_k$ is a basis of $\Hh'$.
Note that the $e_k'$ are indeed unique up to a unitary
transformation, so we only have to show that for one choice of
$e_k'$ we get a unique $T$.
From the equation $\rho=\sigma\circ(\id_\Hh\otimes T)$ we can then
read off the matrix elements of $T$:
\begin{equation}\label{dualT}
  \bra e'_k,\; T(\ketbra{e_\mu}{e_\nu})\;e'_\ell>
    =r_k^{-1/2}r_\ell^{-1/2} \rho\bigl(
        \ketbra{e_k\otimes e_\mu}{e_\ell\otimes e_\nu}\;.
\end{equation}
We have to show that $T$ as defined by this equation is completely
positive whenever $\rho$ is positive. For fixed coefficients $r_k$
the map $\rho\mapsto T$ is obviously linear. Hence it suffices to
prove complete positivity for $\rho=\ketbra\varphi\varphi$. But in
that case $T=V^*AV$ with
$\bra e_\nu,\;Ve_\ell'>
   =r_\ell^{-1/2}\bra e_\ell\otimes e_\nu,\varphi>$, so $T$ is indeed
completely positive. Normalization $T(\idty)=\idty$ follows from
the choice of $r_k$, and the Lemma is proved.

The main use of this Lemma is to translate results about entangled
states to results about channels and conversely. For this it is
necessary to have a translation table of properties. Some entries
are easy: for example, $\rho$ is a product state iff $T$ is
\itx{depolarizing} in the sense that $T(A)=\tr(\rho_2A)$ for some
density operator $\rho_2$, and $\rho$ is separable in the sense of
Definition~\ref{d:separable} iff $T$ is \inx{separable}
(see equation~(\ref{class.telepo})).

\section{\inx{Channel Capacity}}

In the definition of channel capacity we will have to use a
criterion for the approximation of one channel by another. Since
channels are maps between normed spaces, one obvious choice would
be using the standard norm
\begin{equation} \label{norm1}
 \norm{S-T}:=\sup\bigl\lbrace \norm{S(A)-T(A)}\;\mid\;
                  \norm{A}\leq1  \bigr\rbrace\;.
\end{equation}
However, as in the case of positivity there is a problem with this
definition, when one considers tensor products: the norms
$\norm{T\otimes\id_n}$, with $\id_n$ the identity on $\Qm n$, may
increase with $n$. This introduces complications when one has to
make estimates for parallel channels. Therefore we stabilize the
norm with respect to tensoring with ``innocent bystanders'', and
introduce, for any linear map $T$ between C*-algebras
\begin{equation}\label{cbnorm}
   \cbnorm{T}:=\sup_n\norm{T\otimes\id_n}\;,
\end{equation}
called the \itx{norm of complete boundedness}, or ``\inx{cb-norm}'' for
short. This name derives from the observation that on infinite
dimensional C*-algebras the above supremum may be infinite even
though each term in the supremum is finite. By definition, a
completely bounded map is one with $\cbnorm{T}<\infty$. On a
finite dimensional C*-algebra, every linear map is completely
bounded: for maps into $\Qm d$ we have $\cbnorm{T}\leq d\norm{T}$.
(As a general reference for these matters I recommend the book
\cite{Paulsen}). One might conclude from this that the whole
distinction between these norms is irrelevant. However, since we
will need estimates for large tensor products, every factor
increasing with dimension can make a decisive difference. This is
the reason for employing the cb-norm in the definition of channel
capacity. It will turn out, however, that in the most important
cases one has only to estimate differences to the identity, and
$\norm{T-\id}$ and $\cbnorm{T-\id}$ can be estimated in terms of
each other with dimension-independent bounds.

The basis of the notion of channel capacity is the comparison
between the given channel $T:\A_2\to\A_1$ and an ``ideal'' channel
$S:\B_1\to\B_2$. The comparison is affected by suitable encoding and
decoding transformations $E:\A_1\to\B_1$ and $D:\B_2\to\A_2$ so that
the composed operator $ETD:\B_2\to\B_1$ is a map which can be
compared directly with the ideal channel $S$. Of course, we are only
interested in the comparison in the case of optimal encoding and
decoding, i.e., in the quantity
\begin{equation}\label{err}
  \err(S,T)=\inf_{E,D}\cbnorm{S-ETD}\;,
\end{equation}
where the infimum is over all channels (i.e., all unit preserving
completely positive maps) $E$ and $D$ with appropriate domain and
range. Since these data are at least implicitly given together
with $S$ and $T$, there is no need to specify them in the
notation. $S$ should be thought of as representing one word of the
kind of message to be sent, whereas $T$ represents one invocation
of the channel. Channel capacity is defined as the number of
$S$-words per invocation of the channel $T$, which can be
faithfully transmitted with suitable encoding and decoding for
long messages. Here ``messages of length $n$'' are represented by
the tensor power $S^{\otimes n}$, and ``$m$ invocations of the
channel $T$'' are represented by the tensor power $T^{\otimes m}$.

\begin{definition}\label{d.capty}
Let $S$ and $T$ be channels. Then a number $c\geq0$ is called an
``{\bf achievable rate} for $T$ with respect to $S$'', if for any
sequences $n_{\alpha},m_\alpha$ of integers with $m_\alpha\to\infty$
and $\limsup_\alpha(n_\alpha/m_\alpha)< c$ we have
$$ \lim_\alpha \err\bigl(
        S^{\otimes n_\alpha},T^{\otimes m_\alpha}\bigr)=0
\quad.$$
The supremum of all achievable rates is called the {\bf capacity} of
$T$ with respect to $S$, and is denoted by $C(S,T)$.
\end{definition}

Note that by definition $0$ is an achievable rate (no integer
sequences with asymptotically negative ratio exist), hence
$C(S,T)\geq0$. If all $c\geq0$ are achievable, then of course we
write $C(S,T)=\infty$. It may seem cumbersome to check {\it all}
pairs of integer sequences with given upper ratio when testing $c$.
However, due to some monotonicity of $\err$ it suffices to check
only one sequence, provided it is not too sparse: if there is any pair
of sequences $n_\alpha,m_\alpha$ satisfying the conditions in the
definition (including $\err\to0$) plus the extra requirement that
$(m_\alpha/m_{\alpha+1})\to1$, then $c$ is achievable.

The \itx{ideal channel} for systems with observable algebra $\A$ is
by definition the identity map $\id_\A$ on $\A$. For typographical
convenience we will abbreviate ``$\id_\A$'' to ``$\A$'', whenever it
appears as an argument of $\err$ or $\capty$. Using this notation,
we will now summarize the capacities of ideal quantum and classical
channels. Of course, these are basic data for the whole theory.
\begin{eqnarray}
   \capty(\Qm k,\Cl n)&=&0  \;,\quad\hbox{for $k\geq2$}
\label{c:QC}\\
   \capty(\Cl k,\Cl n)
     &=&\capty(\Qm k,\Qm n)
      = \capty(\Qm k,\Cl n)
      = \frac{\log n}{\log k}\;.
\label{c:CC,QQ,CQ}
\end{eqnarray}
Here the first equation is the capacity version of the
No-Teleportation Theorem: It is impossible to transport any
quantum information on a classical channel. The second line shows
that for capacity purposes the $\M_n$ is indeed best compared with
$\C_n$. In classical information theory one uses the 1 bit system $\C_2$
as the ideal reference channel. Similarly, we use the 1 qubit
channel as the reference standard for quantum information , i.e.,
we define the \itx{classical capacity} $\captyc(T)$,
and the  \itx{quantum capacity} $\captyq(T)$, of an arbitrary channel
by
\begin{eqnarray}\label{cap-cq}
     \captyc(T)&=&\capty(\Cl2,T)\\
     \captyq(T)&=&\capty(\Qm2,T)\;.
\end{eqnarray}
Combining the results (\ref{c:CC,QQ,CQ}) with the ``triangle inequality'', or
\itx{two step coding inequality}
\begin{equation}\label{c:chain}
  \capty(T_1,T_3)\geq\capty(T_1,T_2)\capty(T_2,T_3)
\end{equation}
we see that this is really only a choice of \inx{units}, i.e., for
arbitrary channels $T$ we get $\capty(\Qm n,T)=\frac{\log2}{\log
n}\capty(\Qm 2,T)$, and a similar equation for classical
capacities. Note that the term ``\inx{qubit}'' refers to the reference
system $\M_2$, but it is not advisable to use it  as a special
unit for quantum information (rather than just ``bit''): this
would be like distinguishing the units ``vertical meter'' and
``horizontal meter'', and create problems in every equation in
which the two capacities are directly compared. The simplest
relation of this kind is
\begin{equation}\label{Cq<Cc}
  \captyq(T)\leq\captyc(T)\;,
\end{equation}
which follows by combining (\ref{c:chain}) with
(\ref{c:CC,QQ,CQ}).
Note that both definitions apply to arbitrary channels $T$, whether
input and/or output are classical or quantum or hybrids.  In order
for a channel to have positive quantum capacity, it is necessary
that both the input and the output are quantum systems. This is
shown combining (\ref{c:QC}) with the \itx{bottleneck inequality}
\begin{equation}\label{bottle}
  \capty(S,T_1T_2)\leq
     \min\bigl\{\capty(S,T_1),\capty(S,T_2)\bigr\}\;.
\end{equation}
Another application of the bottleneck inequality is to
\itx{separable channels}. These are by definition the channels
with a purely classical intermediate stage, i.e., $T=SR$ with
`output of $S$'= `input of $R$' a classical system. For such
channels $\captyq(T)=0$.

An important operation on channels is running two channels in
parallel, represented mathematically by the tensor product. The
relevant inequality is
\begin{equation}\label{subaddC}
  \capty(S,T_1\otimes T_2)\geq \capty(S,T_1)+ \capty(S,T_2)
\end{equation}
For the standard ideal channels, and when all systems involved are
classical, we even have equality. However, it is one of the big
open problems to decide under what general circumstances this is
true.

\subsubsection{Comparison with the classical definition}

Since the definition of classical capacity $\captyc(T)$ also
applies to the purely {\it classical situation}, we have to verify
that it is indeed equivalent to the standard definition in this
case. To that end we have to evaluate the error quantity
$\cbnorm{T-\id}$ for a classical to classical channel. As noted in
a classical channel $T:\C(Y)\to\C(X)$ is given by a transition
probability matrix $T(x\to y)$. Since the cb-norm coincides with
the ordinary norm in the classical case, we get
\begin{eqnarray*}
  \cbnorm{\id-T}
    &=&\Vert{\id-T}\Vert
     =\sup_{x,f}\Bigl\vert\sum_y\Bigl(\delta_{xy}-T(x\to y)\Bigr)f(y)
               \Bigr\vert \\
    &=&2\sup_x \bigl(1-T(x\to x)\bigr)\;
\end{eqnarray*}
where the supremum is over all $f\in\C(Y)$ with $\abs{f(y)}\leq1$
and is attained where $f$ is just the sign of the parenthesis in
the second line, and we used the normalization of transition
probabilities. Hence, apart from an irrelevant factor two,
$\cbnorm{T-\id}$ is just the \itx{maximal probability of error},
i.e., the largest probability for sending $x$ and getting anything
different. This is precisely the quantity, which is demanded to go
to zero (after suitable coding and decoding) in Shannon's
classical definition of the channel capacity of discrete
memoryless channels \cite{Shannon}. Hence the above definition
agrees with the classical one.

When considering the classical capacity $\captyc(T)$ of a quantum
channel, it is natural to look at a coded channel $ETD$, as a
channel in its own right. Since we consider transmission of
classical information, this is a purely classical channel, and we
can look at its classical capacity. Optimizing over coding and
decoding, we get the quantity
\begin{equation}\label{oneshotcc}
  C_{c,1}(T)=\sup_{ETD\ \rm classical} \captyc(ETD)\;.
\end{equation}
This is called the \itx{one-shot classical capacity}, because it
seems to involve only one invocation of the channel $T$. Of
course, many uses of the channel are implicit in the capacity on
the right hand side, but these are in some sense harmless. In
fact, every coding and decoding scheme for comparing
$(ETD)^{\otimes n}$ to an ideal classical channel is also a
coding/decoding for $T^{\otimes n}$, but the codings/decodings
arising in this way from coding $ETD$ are only those, in which the
coded input states and measurements at the outputs are {\it not
entangled}. If we allow entanglement over blocks of a large length
$\ell$ we thus recover the full classical capacity:
\begin{equation}\label{1shot-nshot}
     C_{c,1}(T)\leq C_c(T)
               =\sup_\ell\ \frac1\ell\ C_{c,1}(T^{\otimes\ell})\;.
\end{equation}
It is not clear, whether equality holds here. This is a
fundamental question, which can be paraphrased as this: ``Does
\inx{entangled coding} ever help for sending classical information over
quantum channels?''. At the moment all partial results known to me
seem to say that this is not the case.

\subsubsection{Comparison with other error criteria}
Coming now to the quantum capacity $\captyq(T)$, we have relate
our definition to more current ones. One version, first stated by
Bennett is very similar to the one given above, but differs
slightly in the error quantity, which is required to go to zero.
Rather than $\cbnorm{T-\id}$, he considers the lowest
\itx{fidelity} of the channel, defined as
\begin{equation}\label{fidelCap}
\fidel(T)=\inf_{\psi}
     \langle\psi,T\bigl(\ketbra\psi\psi\bigr)\psi\rangle\;,
\end{equation}
where the supremum is over all unit vectors. Hence achievable
rates are those for which $\fidel(ET^{\otimes n_\alpha}D)\to1$,
where $E,D$ map to a system of $m_\alpha$ qubits, and these
integer sequences satisfy the same constraints as above. This is
definition is equivalent to ours, because the error estimates are
equivalent. In fact, if we introduce the \itx{off-diagonal
fidelity}
\begin{equation}\label{offdifi}
  \fidel_\%(T)=\sup_{\phi,\psi}
              \Re\!e\bra\phi,T\bigl(\ketbra\phi\psi\bigr)\psi>
\end{equation}
for any channel $T:\Qm d\to\Qm d$ with $d<\infty$, we have the
following system of estimates:
\begin{eqnarray}
  \norm{T-\id} &\leq& \cbnorm{T-\id}
               \leq 4\sqrt{1-\fidel_\%(T)}
               \leq 4\sqrt{\norm{T-\id}}\\
  \norm{T-\id} &\leq& 4\sqrt{1-\fidel(T)}
               \leq 4\sqrt{1-\fidel_\%(T)}  \;,
\end{eqnarray}
which will be proved elsewhere. The main point is, though that the
dimension does not appear in these estimates, so if one such
quantity goes to zero, all others do, and we can build an
equivalent capacity definition on any one of them.

Yet another definition of quantum capacity has been given in terms
of entropy quantities \cite{schum}, and was also shown to be equivalent
\cite{BarNilSen}.

\section{Coding Theorems}

The definition of channel capacity looks simple enough, but
computing it on the basis of this definition is in general a very
hard task: it involves an optimization over all coding and
decoding channels in systems of asymptotically many tensor
factors. Hence it is crucial to get simpler expressions, which can
be computed in a much more direct way from the matrix elements of
the given channel. Such results are called \itx{coding theorems},
after the first theorem of this type, established by Shannon.

In order to state it we need some entropy quantities. The \itx{von
Neumann entropy} of a state with density matrix $\rho$ is defined
as
\begin{equation}\label{vNeumann}
  S(\rho)=-\tr\bigl(\rho\log\rho\bigr)\;,
\end{equation}
where the function of $\rho$ is evaluated in the functional
calculus, and $0\log0$ is defined to be zero. The logarithm will
be chosen to be base 2, so the unit for entropies is ``{\tt
bit}''. The \itx{relative entropy} of a state $\rho$ with respect
to another, $\sigma$, is defined by
\begin{equation}\label{relEnt}
  S(\rho,\sigma)=\tr\bigl(\rho(\log\rho-\log\sigma)\bigr);
\end{equation}
Both quantities are positive, and may be infinite on an infinite
dimensional space. The von Neumann entropy is concave, whereas the
relative entropy is convex jointly in both arguments. For more
precise definitions, and many further results, I recommend the
book of Petz and Ohya~\cite{PetzOh}.

The strongest coding theorem for quantum channels known so far is
the following expression for the one-shot classical capacity,
proved by Holevo \cite{holevoCT}
\begin{equation}\label{CcodeT}
   C_{c,1}(T)=\max \left[
              S\Bigl( \sum_{i}p_{i}T_*[\rho_{i}]\Bigr)
              - \sum_{i}p_{i}S(T_*[\rho_{i}])  \right]
\end{equation}
Whether or not this is equal to the  classical capacity depends on
whether the conjectured equality in equation~(\ref{1shot-nshot})
holds or not. In any case, it is known to hold for channels with
classical input, so \inx{Holevo's coding theorem} is a genuine extension
of Shannon's.

For the quantum capacity no coding theorem has been proved yet.
However, there is a fairly good candidate for the right hand side,
related to a quantity called ``\itx{coherent information}''
\cite{cohinf}. The formula is written most compactly by relating
it to an entanglement quantity via Lemma~\ref{l.dual}. For any
bipartite state $\rho$ with restriction $\rho^B$ to the second
factor, let
\begin{equation}\label{cerf}
  E_S(\rho)=S(\rho^B)-S(\rho)\;.
\end{equation}
This is an entanglement measure of sorts, because it is large when
$S(\rho)$ is small, e.g., when $\rho$ is pure, and $\rho^B$ is
very mixed, e.g., when $\rho$ is maximally entangled. It can be
negative, though (see \cite{Cerf} for a discussion). Then we set
\begin{equation}\label{QcodeT1}
  C_{S,1}(T)=\sup_\sigma\;E_S(\sigma\circ(\id\otimes T))\;,
\end{equation}
where the supremum is over all bipartite pure states $\sigma$.
Note that any measure of entanglement can be turned into a
capacity-like expression by this procedure. Since this quantity is
known not to be additive \cite{VSS}, the candidate for the right
hand side of the quantum coding theorem is
\begin{equation}\label{QcodeT}
  C_{S}(T)=\sup_\ell\ \frac1\ell\ C_{S,1}(T^{\otimes\ell})\;,
\end{equation}
in analogy to (\ref{1shot-nshot}). So far there are some good
heuristic arguments \cite{lloyd,horochan} in that direction, but a
full proof remains one of the main challenges in the field.

An interesting upper bound on $\captyq(T)$ can be written in terms
of the \inx{transpose} operation $\Theta$ on the output system
\cite{HolWer}: one has
\begin{equation}\label{QcodeEst}
   \captyq(T)\leq\log_2\cbnorm{\Theta T}\;.
\end{equation}
Hence if $\Theta T$ happens to be completely positive (as for any
channel with an intermediate classical state) this map is a
channel, hence has cb-norm $1$, and $\captyq(T)=0$. This criterion
can also be used to show that whenever there is sufficiently high
noise in a channel, it will have quantum capacity zero.

\section{Teleportation and Dense Coding Schemes\label{sec:teleposd}}

In this section we will show that entanglement assisted
teleportation and dense coding as described in
Sections~\ref{sec:telepo} and \ref{sec:sdcode} really work.

Rather than going through the now standard derivations in the
basic qubit examples, we will use the structure assembled so far
to reverse the question, i.e., we try to find the {\it most
general} setup in which teleportation and dense coding work
without errors. This will some give additional insights, and
possibly some welcome flexibility when it comes to realizing these
processes for larger than qubit systems. The task as stated is
somewhat beyond the scope of this paper, mainly because there are
so many ways to waste resources, which do not necessarily have a
compact characterization. So in order to get a readable result, we
only look at the ``tight case'' \cite{alltelepo}, in which
resources are used in a sense optimally. By this we mean that all
Hilbert spaces involved have the same finite but arbitrary
dimension $d$ (so we can take them all equal to $\Hh=\Cx^d$), and
the classical channel distinguishes exactly $\abs X=d^2$ signals.

For both teleportation and dense coding the beginning of each
transmission is to distribute the parts of an entangled state
$\omega$ between sender Alice and receiver Bob. Only then Alice is
given the message she is supposed to send, which is a quantum
state in the case of teleportation and a classical value in case
of dense coding. She codes this in a suitable way, and Bob
reconstructs the original message by evaluating Alice's signal
jointly with his entangled subsystem.

For {\em \inx{dense coding}}, assume that $x\in X$ is the message given
to Alice. She encodes it by transforming her entangled system by a
channel $T_x$, and sending the resulting quantum system to Bob,
who measures an observable $F$ jointly on Alice's particle and
his. The probability for getting $y$ as a result is then
$\tr\bigl(\omega (T_x\otimes\id)(F_y)\bigr)$, where the
``$\otimes\id$'' expresses the fact that no transformation is done
to Bob's particle while Alice applies $T_x$ to hers. If everything
works correctly, this expression has to be $1$ for $x=y$, and $0$
otherwise:
\begin{equation}\label{densecod}
  \tr\bigl(\omega (T_x\otimes\id)(F_y)\bigr)=\delta_{xy}\;.
\end{equation}

Let us take a similar look at {\it \inx{teleportation}}. Here three
quantum systems are involved: the entangled pair in state
$\omega$, and the input system given to Alice, in state $\rho$.
Thus the overall initial state is $\rho\otimes\omega$. Alice
measures an observable $F$ on the first two factors, obtaining a
result $x$ sent to Bob. Bob applies a transformation $T_x$ to his
particle, and makes a final measurement of an observable $A$ of
his choice. Thus the probability for Alice measuring $x$ and for
Bob getting a result ``yes'' on $A$, is
$\tr(\rho\otimes\omega)(F_x\otimes T_x(A))$. Note that the tensor
symbols in this equation refer to different splittings of the
system ($1\otimes23$ and $12\otimes3$, respectively).
Teleportation is successful, if the overall probability for
getting $A$, computed by summing over all possibilities $x$, is
the same as for an ideal channel, i.e.,
\begin{equation}\label{telepo}
  \sum_{x\in X}\tr(\rho\otimes\omega)(F_x\otimes T_x(A))
   =\tr(\rho A) \;.
\end{equation}

Surprisingly, in the tight case one gets exactly the same
conditions on $\omega,T_x,F_x$ for teleportation and dense coding,
i.e., a dense coding scheme can be turned into a teleportation
scheme simply by letting Bob and Alice swap their equipment.
However, this symmetry depends crucially on the tightness
condition, because teleportation schemes with $\abs X>d^2$ signals
are trivial to get, but $\abs X>d^2$ is impossible for dense
coding. Conversely, dense coding through a $d'>d$ dimensional
channel is trivial to get, while teleportation of states with
$d'>d$ dimensions (with the same $X$) is impossible.

Let us now give a heuristic sketch of the arguments leading to the
necessary and sufficient conditions on  for equations
(\ref{telepo}) and (\ref{densecod}) to hold. For full proofs we
refer to \cite{alltelepo}. A crucial ingredient for the analysis
of the teleportation equation is the ``No measurement without
perturbation'' principle from Lemma~\ref{niwp}: the left hand side
of (\ref{telepo}) is indeed such a decomposition, so each term
must be equal to $\lambda_x\tr(\rho A)$ for all $\rho, A$. But we
can carry this even further: suppose we decompose $\omega,F_x$, or
$T_x$ into a sum of (completely) positive terms. Then each term in
the resulting sum must also be proportional to $\tr(\rho A)$.
Hence any components of $\omega,T_x$ and $F_x$ satisfy a
teleportation equation as well (up to normalization). Similarly,
the vanishing of the dense coding equation for $x\neq y$ carries
over to every positive summand in $\omega,T_x$, or $F_x$. Hence it
is plausible that we must first analyze the case where all
$\omega,F_x,T_x$ are ``pure'', i.e., have no non-trivial
decompositions as sums of (completely) positive terms:
\begin{eqnarray}\label{tp:pure}
  \omega&=&\ketbra\Omega\Omega\\
  F_x   &=&\ketbra{\Phi_x}{\Phi_x}\\
  T_x(A)&=&U_x^*AU_x\;.\label{tp:pureT}
\end{eqnarray}
The further analysis will show that in the pure case any two of
these elements determine the third via the teleportation or the
dense coding equation, so that in fact all components of $\omega$
(resp. $T_x$ or $F_x$) have to be proportional. Hence each of
these has to be pure in the first place. For the present
discussion, let us just assume purity in the form
(\ref{tp:pure},...,\ref{tp:pureT}) from now on. Note that
normalization requires that each $U_x$ is unitary.

The second normalization condition,
$\sum_x\ketbra{\Phi_x}{\Phi_x}=\sum_xF_x=\idty$ has an interesting
consequence in conjunction with the tightness condition: the
vectors $\Phi_x$ live in a $d^2$-dimensional space, and there are
exactly $d^2$ of them. This implies that they are orthogonal:
Since each vector $\Phi_x$ satisfies $\norm{\Phi_x}\leq1$, and
$d=\tr(\idty)=\sum_x\norm{\Phi_x}^2$, we must have
$\norm{\Phi_x}=1$ for all $x$. Hence in the sum
$1=\sum_x\bra\Phi_y,F_x\Phi_y>$ the term $y=x$ is equal to $1$,
and hence the others must be be zero.

Now consider the term with index $x$ in the teleportation equation
and set $\rho=\ketbra{\phi'}{\phi}$ and $A=\ketbra{\psi}{\psi'}$.
Then the trace splits into two scalar products, in which the
variables ${\phi},{\phi'},{\psi},{\psi'}$ can be chosen
independently, which leads to an equation of the form
\begin{equation}\label{halftelepo}
  \bra\phi\otimes\Omega, \Phi_x\otimes(U_x^*\psi)>
  =\lambda_x\bra\phi,\psi>\;,
\end{equation}
for all $\phi,\psi$, and coefficients which must satisfy
$\sum_x\abs{\lambda_x}^2=1$. Note how in this equation a scalar
product between the vectors in the first and third tensor factor
is generated. This type of equation, which is clearly the core of
the teleportation process may be solved in general:

\begin{Lem}Let $\Hh,\K$ be finite dimensional Hilbert spaces, and
let $\Omega_1\in\K\otimes\Hh$ and $\Omega_2\in\Hh\otimes\K$ be
unit vectors such that, for all $\phi,\psi\in\Hh$,
\begin{equation}\label{telepo/2}
  \bra\phi\otimes\Omega_1, \Omega_2\otimes\psi>
     =\lambda\bra\phi,\psi>\;.
\end{equation}
Then $\abs\lambda\leq1/\dim\Hh$ with equality iff $\Omega_1$ and
$\Omega_2$ are maximally entangled and equal up to the exchange of
the tensor factors $\Hh$ and $\K$.
\end{Lem}

For the proof consider the Schmidt decomposition
$\Omega_1=\sum_k\sqrt{w_k} f_k\otimes e_k$, and insert $\phi=e_n$,
$\psi=e_m$ into equation~(\ref{telepo/2}) to find the matrix
elements of $\Omega_2$:
\begin{displaymath}
 \bra e_n\otimes f_m, \Omega_2>
    =\lambda\; w_m^{-1/2}\ \delta_{nm}\;.
\end{displaymath}
Clearly, $\norm{\Omega_2}^2=\abs\lambda^2\sum_mw_m^{-1}$. This sum
takes its smallest value under the constraint
$\sum_mw_m=\norm{\Omega_1}^2=1$ only at the point where all $w_m$
are equal. This proves the Lemma.

We apply it to $\Omega_1=(\idty\otimes U_x)\Omega$, and
$\Omega=\Phi_x$. Then $\sum_x\abs{\lambda_x}^2\leq d^2d^{-2}=1$,
with equality only if all the vectors involved are maximally
entangled and pairwise equal up to an exchange of factors:
\begin{equation}\label{Phi2U}
  \Phi_x=(U_x\otimes\idty)\Omega\;,
\end{equation}
where we take $\Omega=d^{-1/2}\sum_ke_k\otimes e_k$ by an
appropriate choice of bases. If $\Omega$ is maximally entangled,
equation~(\ref{Phi2U}) sets up a one-to-one correspondence between
unitary operators $U_x$ or the vectors $\Phi_x$ as independent
elements in the construction. The $\Phi_x$ have to be an
orthonormal basis of maximally entangled vectors, and there are no
further constraints. In terms of the $U_x$ the orthogonality of
the $\Phi_x$ translates in the orthogonality with respect to the
Hilbert-Schmidt scalar product:
\begin{equation}\label{Uortho}
  \tr(U_x^*U_y)=d\; \delta_{xy}\;.
\end{equation}
Again, there are no further constraints, so any collection of
$d^2$ unitaries satisfying these equations leads to a
teleportation scheme.

For dense coding case we get the same result, although along other
routes. Equation~(\ref{Phi2U}) follows easily by writing the
teleportation equation as
$\abs{\bra\Omega,(U_x^*\otimes\idty)\Phi_x>}^2=\delta_{xy}$. The
problem is to show that $\Omega$ has to be maximally entangled.
Using the reduced density operator $\omega_1$ of $\omega$, this
becomes
\begin{equation}\label{skewUortho}
 \tr(\omega_1 U_x^*U_y)
 =\bra\Omega, (U_x^*U_y\otimes\idty)\Omega>
 =\bra\Phi_x,\Phi_y>
 =\delta_{xy}\;.
\end{equation}
We claim that this equation, for a positive operator $\omega_1$,
and $d^2$ unitaries $U_x$, implies that $\omega_1=d^{-1}\idty$. To
see this, expand the operator $A=\ketbra\phi{e_k}\omega_1^{-1}$ in
the basis $U_x$ according to the formula
$A=\sum_xU_x\tr(U_x^*A\omega_1)$:
\begin{displaymath}
  \sum_x \bra e_k,U_x^*\phi>\ U_x=\ketbra{\phi}{e_k}\omega_1^{-1}\;.
\end{displaymath}
Taking the matrix element $\langle\phi\vert\cdot\vert e_k\rangle$
of this equation, and summing over $k$, we find
\begin{displaymath}
  \sum_{x,k} \bra e_k,U_x^*\phi>\ \bra \phi,U_xe_k>
  =\sum_x\tr(U_x^*\ketbra\phi\phi U_x)
  =d^2\norm\phi^2
     =\norm\phi^2\; \tr(\omega_1^{-1})\;.
\end{displaymath}
Hence $\tr(\omega_1^{-1})=d^2=\sum_kr_k^{-1}$, where $r_k$ are the
eigenvalues of $\omega_1$. Using again that the smallest value of
this sum under the constraint $\sum_kr_k=1$ is attained only for
constant $r_k$, we find $\omega_1=d^{-1}\idty$, and $\Omega$ is
indeed maximally entangled.

To summarize, we have the following Theorem (again, for a detailed
proof see \cite{alltelepo}):

\begin{theorem}Given either a teleportation scheme or a dense
coding scheme, which is tight in the sense that all Hilbert spaces
are $d$-dimensional, and $\abs X=d^2$ classical signals are
distinguished. Then
\begin{itemize}
\item $\omega=\ketbra\Omega\Omega$ is pure and maximally
entangled,
\item $F_x=\ketbra{\Phi_x}{\Phi_x}$, where the $\Phi_x$ form an
orthonormal basis of maximally entangled vectors,
\item $T_x(A)=U_x^*AU_x$, where the $U_x$ are unitary and
orthonormal in the sense that $\tr(U_x^*U_y)=d\;\delta_{xy}$, and
\item these objects are connected by the equation
$\Phi_x=(U_x\otimes\idty)\Omega$.
\end{itemize}
Given either the $\Phi_x$ or the $U_x$ with the appropriate
orthogonality properties, and a maximally entangled vector
$\Omega$, the above conditions determine a dense coding and a
teleportation scheme.
\end{theorem}

In particular, we have shown that a teleportation scheme becomes a
dense coding scheme, and conversely, when Alice and Bob swap their
equipment. However, this is only true in the tight case: for a
larger quantum channel dense coding becomes easier but teleporting
becomes more demanding. Similarly, teleportation becomes easier
with more allowed classical information exchange, whereas dense
coding of more than $d^2$ signals is impossible.

In order to construct a scheme, it is best to start from the
equation $\tr(U_x^*U_y)=d\;\delta_{xy}$, i.e., to look for
orthonormal bases in the space of operators consisting of
unitaries. For $d=2$ the solution is essentially unique:
$U_1,\ldots,U_4$ are the identity and the three Pauli matrices,
which leads to the standard examples of. Group theory helps to
construct examples of such bases for any dimension $d$, but this
construction by no means exhausts the possibilities. A fairly
general construction is given in \cite{alltelepo}. It requires two
combinatorial structures known from classical \itx{design theory}
\cite{BethJ}: a \itx{Latin square} of order $d$, i.e., a matrix in
which each row and column is a permutation of $(1,...,d)$, and $d$
\itx{Hadamard matrices}, i.e., unitary $d\times d$-matrices, in
which each entry has modulus $d^{-1/2}$. For neither Latin squares
nor Hadamard matrices an exhaustive construction exists, so these
are rich fields for hunting and gathering new examples, or even
infinite families of examples. Certainly this connection suggests
that a full classification or exhaustive construction of
teleportation and dense coding schemes cannot be expected.
However, it may still be a good project to look for schemes with
additional desirable features.

\end{document}